\documentclass[9pt,twocolumn,twoside]{pnas-new}

\templatetype{pnasmathematics} 

\setboolean{displaywatermark}{false}
\usepackage{physics}
\usepackage{amsmath}
\usepackage{caption}
\usepackage{subcaption}

\title{Duality of the Principle of Least Action: A New Formulation of Classical Mechanics}

\author[a]{David J. Tannor}

\affil[a]{ Weizmann Institute of Science, Rehovot 76100, Israel}

\leadauthor{David J. Tannor}

\significancestatement{This article provides a new and simple derivation of the key equations of analytical mechanics, with no assumption other than the convexity of the action.}

\authorcontributions{}
\authordeclaration{The authors declare no conflict of interests.}
\correspondingauthor{\textsuperscript{1} E-mail: david.tannor@weizmann.ac.il}


\begin{abstract}
A dual formalism for Lagrange multipliers is developed. The formalism is used to minimize an action function $S(q_2,q_1,T)$ without any dynamical input other than that $S$ is convex.  All the key equations of analytical mechanics – the Hamilton-Jacobi equation, the generating functions for canonical transformations, Hamilton’s equations of motion and $S$ as the time integral of the Lagrangian – emerge as simple consequences. It appears that to a large extent, analytical mechanics is simply a footnote to the most basic problem in the calculus of variations: that the shortest distance between two points is a straight line.  
\end{abstract}

\dates{This manuscript was compiled on \today}
\doi{\url{www.pnas.org/cgi/doi/10.1073/pnas.XXXXXXXXXX}}

\begin{document}

\maketitle
\thispagestyle{firststyle}
\ifthenelse{\boolean{shortarticle}}{\ifthenelse{\boolean{singlecolumn}}{\abscontentformatted}{\abscontent}}{}

\section{Introduction}
\label{sec:introduction}
\dropcap{T}his article has three key ideas. 1) A doubling of the theory of Lagrange multipliers, what we call a dual theory.  In colloquial language, the dual formalism views the Lagrange multiplier as a way of violating a constraint with the help of an undetermined multiplier. At the end, the violation is corrected by choosing the value of the undetermined multiplier to satisfy the constraint.  For every minimization problem, the dual formalism leads to a maximization problem and vice versa. 2) The dual Lagrange formalism is applied first to the entropy maximum principle in thermodynamics and then to the principle of least action in mechanics.  In the former, the total entropy of two subsystems is maximized with the constraint that the total energy is conserved. In the latter, the action $S$ is minimized over two segments of a path connecting $q_1$ to $Q$ and $Q$ to $q_2$, subject to the constraint that $t_1+t_2=T$, i.e. the total time is constrained. The analogy with the entropy maximum principle is clear: in one case we maximize over all partitionings of energy between two subsystems, in the other we minimize the action over all partitionings of time over the two segments of the path. Just as in the former, the only assumption is concavity of the entropy, in the latter the only assumption is convexity of the action – there is no further dynamical input. 3) We then continue to apply the dual formalism to mechanics, now allowing the endpoints $q_1$, $q_2$ and the intermediate point $Q$ to vary. Viewing the path of least action from $q_1$ to $q_2$ in time $T$ as finding the shortest distance from $q_1$ to $q_2$ in a Riemannian space, we find three ways to violate the constraints: a) allowing more than the allotted time; b) moving the endpoints closer; c) allowing the end of segment one, $Q_1$, and the beginning of segment two, $Q_2$, to not be identical.  The formalism gives as simple consequences: a) the Hamilton-Jacobi equation and its dual form for the abbreviated action $\bar{S}(E)$; b) the $F_1$ generating function relations for canonical transformations and the $F_2-F_4$ generating functions from dual forms; c) Hamilton’s equations of motion, with the two Hamilton equations of motion emerging from the standard and the dual Lagrange formalism,  and d) the action $S(q_1,q_2,T)$ as the time integral of the Lagrangian and its dual form $\bar{S}(q_1,q_2,E)$ as the integral of the 1-form $pdq$. It is noteworthy that these latter two equation emerge at the end our development, as opposed to conventional treatments when they appear at the beginning.

The approach provides a simple and integrated approach to the derivation of the key equations of analytical mechanics. Moreover, it clears up numerous conceptual difficulties in the standard presentations of this material, some salient examples of which are given in the Conclusions. Perhaps most importantly, the approach reveals that to a large extent, analytical mechanics is simply a footnote to a mathematical problem in the calculus of variations: that the shortest distance between two points is a straight line. 

A note about the title of the paper. The phrase "duality of the principle of least action" is used in two senses.  The first is the liberal use of the term to refer to a novel formulation of the principle of least action in terms of the dichotomous partitioning of the total time and distance into two segments. This dichotomous partitioning, when combined with Lagrange multipliers, is responsible for the primary results of the paper. The second sense is the combining of the principle of least action with the dual Lagrange multiplier formalism; this is the sense of duality used by mathematicians and leads to the dual equations associated with each of the primary equations \footnote{For simplicity, the terminology ''principle of least action" is used throughout this paper. In general the principle describes stationary points of the action that are not necessarily minima.}

\section{Lagrange Multipliers and Dual Formalism}
\label{sec:lagrange_intro}
Lagrange multipliers are widely used in mathematics and all branches of physics to solve extremal problems in the presence of constraints. One adds the constraint equation to the function to be minimized using a Lagrange multiplier, and then finds the extremal of the modified problem with $N+1$ variables, where the Lagrange multiplier provides an extra unknown and an extra equation.  There is an alternative, complementary interpretation of Lagrange multipliers.  Adding the constraint to the original function with an undetermined multiplier \emph{deconstrains} the original extremization, so if the original task was to minimize the function, the new minimum without constraints will always be below the true minimum and hence a lower bound to the true minimum.  Maximizing this lower bound with respect to the undetermined multiplier can only approach but never exceed the true minimum and for an important class of problems (including those relevant to thermodynamics and classical mechanics) gives the minimum to the original constrained problem.  We call this a dual Lagrange multiplier formalism.  In colloquial terms, the strategy is to violate the constraints, then solve for the extremum, and finally to choose the multiplier so that the constraints are satisfied.  For every minimization problem, the dual formalism leads to a maximization problem and vice versa. This duality of Lagrange multipliers is not included in most of the standard textbooks on analytical mechanics \cite{goldstein,lanczos}, control theory \cite{bryson} or mathematical methods of physics \cite{arfken}, although there are more advanced books and articles that treat it \cite{hilbert,walsh,bertsekas,kalman}.

\subsection{Lagrange multipliers}
\label{sec:lagrange}
\begin{figure}[h!]
  \begin{center}
\includegraphics[width=15cm]{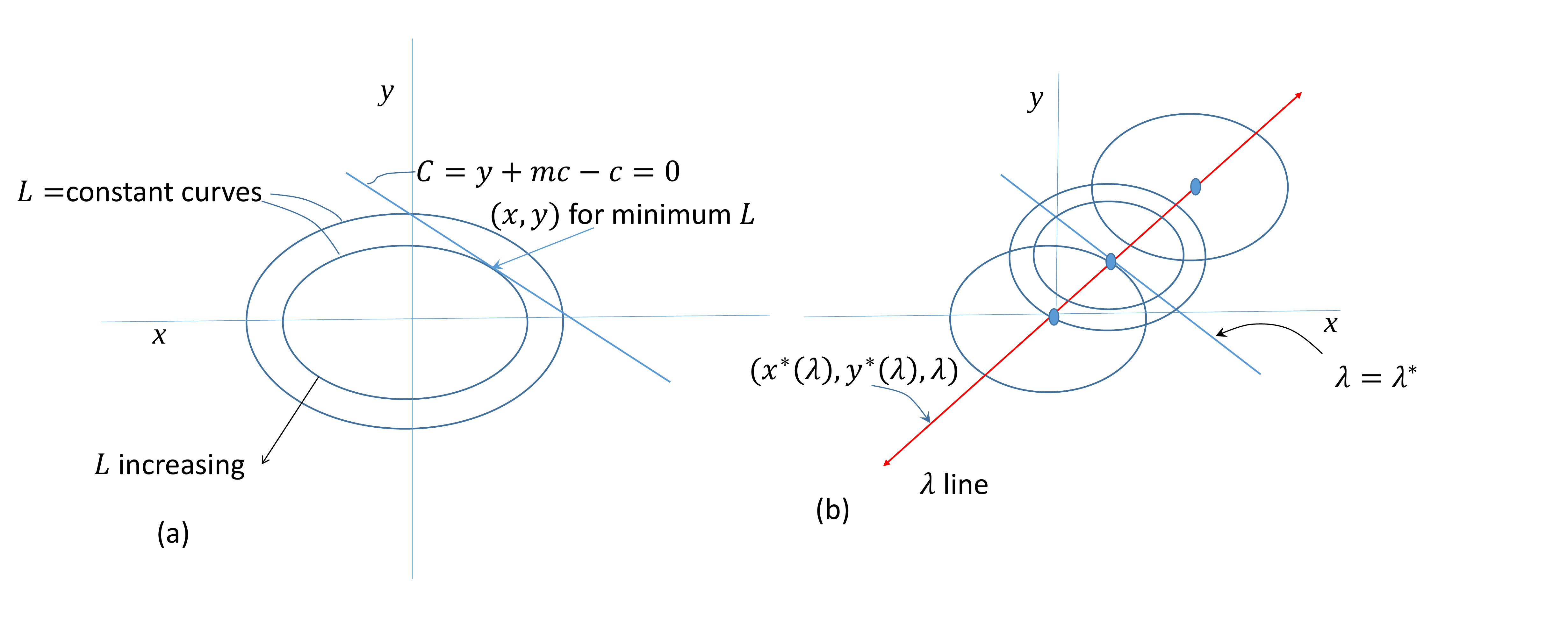}
\end{center}
\caption{a. Standard representation of Lagrange multipliers. b. Dual representation of Lagrange multipliers }
\label{fig:lagrange_ab}
\end{figure}

The following example is based on ref. \cite{bryson}. Consider the problem of minimizing the function 
\begin{equation}
L(x,y) = 1/2 (x^2/a^2 + y^2/b^2)
\end{equation}
subject to the constraint 
\begin{equation}
C(x,y)=y+mx-c=0.
\label{eq:constraint}
\end{equation} In the method of Lagrange multipliers, the constraint equation is added to the function to be extremized with a Lagrange multiplier $\lambda$: 
\begin{equation}
H(x,y,\lambda) = L(x,y) + \lambda C(x,y) = 1/2 (x^2/a^2 + y^2/b^2) + \lambda( y+mx-c).
\label{eq:l+lambda_c}
\end{equation}
Setting the derivatives of $H$ with respect to $x$,$y$ equal to 0 leads to two equations, in addition to the constraint equation, for the three unknowns, $x$,$y$ and $\lambda$. Carrying out the derivative calculation:
\begin{equation}
(\partial H/\partial x)_{y,\lambda}=0 = x/a^2 + \lambda m \Longrightarrow x^*=-\lambda m a^2 = x^*(\lambda)
\label{eq:x_star}
\end{equation}
\begin{equation}
(\partial H/\partial y)_{x,\lambda}=0 = y/b^2 + \lambda \Longrightarrow y^*=-\lambda b^2 = y^*(\lambda),
\label{eq:y_star}
\end{equation}
where we use the notation $x^*$, $y^*$ to indicate the root values of $x$,$y$. Substituting $x^*$ and $y^*$ into eq. \ref{eq:constraint} yields: 
\begin{equation}
\lambda^*=-c/(b^2 +m^2a^2),
\label{eq:lambda*}
\end{equation}
i.e. the undetermined multiplier $\lambda$ is now determined to be $\lambda^*$ to satisfy the constraints. Having found $x^*$, $y^*$ and $\lambda^*$ we may substitute into $H(x,y,\lambda)$ to find the constrained minimum $H(x^*,y^*,\lambda^*)=c^2/(2(m^2a^2 + b^2))=L^*(x^*(\lambda^*),y^*(\lambda^*))$.
The method has a physical interpretation of requiring that the tangent of $\grad L$ and the tangent of $\grad C$ be parallel, i.e. that the extremum is found when the value of $L(x,y)$ is tangent to the constraint (see Fig. \ref{fig:lagrange_ab}a).  In our example, $\grad L=x/a^2 \hat{x} + y/b^2 \hat{y}$ and $\grad C = m \hat{x} + 1 \hat{y}$. Substituting $x^*$ and $y^*$ into the expression for $\grad L$ we find:
$\grad L=-\lambda m \hat{x} -\lambda \hat{y}$, which is proportional to $\grad C$ for any value of $\lambda$.

Alternatively, the Lagrange multiplier method may be viewed as expanding the original problem of two equations for two unknowns, $x$ and $y$ (subject to a constraint), to a new problem of three equations in three unknowns, $x$,$y$ and $\lambda$.  To eqs. \ref{eq:x_star}-\ref{eq:y_star} we add 
\begin{equation}
(\partial H/\partial \lambda)_{x,y} =0 \Longrightarrow y+mx-c=0,
\label{eq:constraint_lambda}
\end{equation}
providing the third equation; note that the third equation just recovers the constraint. 

\subsection{Dual formalism}
\label{sec:lagrange_dual}
In the dual formulation of Lagrange multipliers, we substitute $x^*$ and $y^*$
back into $H$. After simplifying we find:
\begin{equation}
H(x^*(\lambda),y^*(\lambda),\lambda)= -\lambda(\lambda m^2 a^2/2 + \lambda b^2/2 +c),
\end{equation}
where $H$ is now a function of $\lambda$ only. Taking the total derivative of $H$ with respect to $\lambda$ and setting it equal to zero,
\begin{equation}
dH(\lambda)/d \lambda = 0 = -\lambda m^2 a^2 - \lambda b^2 - c,
\label{eq:total_deriv}
\end{equation}
we find:
\begin{equation}
\lambda^*=-c/(b^2 +m^2a^2),
\label{eq:lambda*}
\end{equation}
in agreement with eq. \ref{eq:lambda*} for $\lambda^*$.
Note that the total derivative with respect to $\lambda$ of the dual form $H(x^*,y^*,\lambda)$, eq. \ref{eq:total_deriv}, is equal to the partial derivative with respect to $\lambda$ of the original form $H(x,y,\lambda)$ at $x^*$, $y^*$, eq. \ref{eq:constraint_lambda}, with eqs. \ref{eq:x_star}-\ref{eq:y_star}. This is a general feature of the dual formalism.  It arises since
\begin{equation}
dH(x^*(\lambda),y^*(\lambda),\lambda)/d \lambda = (\partial H/\partial x)_{x^*}(dx^*/d\lambda) 
+ (\partial H/\partial y)_{y^*}(dy^*/d \lambda) + \partial H/\partial \lambda = \partial H/\partial \lambda
\end{equation}
and $(\partial H/\partial x)_{x^*}=0$, $(\partial H/\partial y)_{y^*}=0$ at $x^*$ and $y^*$.  

Calculating the second derivative of $H$ with respect to $\lambda$ we find:
\begin{equation}
d^2H/d \lambda^2 = - ma^2 - b^2 < 0,
\end{equation}
i.e. if $x^*$,$y^*$ provide a minimum of $H(x,y)$,  (i.e. $\partial^2 H/\partial x^2 >0, \partial^2H/\partial y^2>0$), they now provide a maximum of $H(x^*(\lambda),y^*(\lambda),\lambda)$ with respect to $\lambda$ $(d^2H^*/d \lambda^2 <0)$,with the same value $H(x^*(\lambda^*),y^*(\lambda^*),\lambda^*)=c^2/(2(m^2a^2+b^2))$. This mathematical property --- that the original minimization problem becomes a maximization problem with respect to $\lambda$ --- implies that in the extended space of $(x,y,\lambda)$ the solution is actually a saddle point. The appearance of a maximum with respect to $\lambda$ has an intriguing conceptual explanation.  Adding the constraint to the original function \emph{deconstrains} the problem – in effect it allows one to solve the problem as if there were no constraints.  
This has the following consequence. If the original problem is to minimize a function, since the constraints are removed the minimum one finds will be lower than or equal to the true minimum; thus it serves as a lower bound to the true minimum.  Since the lower bound is expressed in terms of $\lambda$, maximizing the lower bound with respect to $\lambda$ one obtains the greatest lower bound.  The result of the maximization cannot exceed the physical minimum, and for a wide class of problems actually produces the physical minimum.

\subsection{Geometrical Formulation} 
\label{sec:geometrical}
There is an interesting geometrical interpretation to the dual formulation. We may rewrite eq. \ref{eq:l+lambda_c} as
\begin{equation}
H=H''+C''
\label{eq:h"c"}
\end{equation}
where
\begin{equation}
H''=1/2 [(x/a + m \lambda a)^2 + (y/b + \lambda b)^2],~~~~~~C''=- 1/2 m^2 \lambda^2 a^2 - 1/2 \lambda^2 b^2 - \lambda c=C''(\lambda).
\label{eq:c''}
\end{equation}
Equations \ref{eq:h"c"}- \ref{eq:c''} indicate that the contours of $H(x,y, \lambda)$ are ellipses with a minimum (with respect to $x$ and $y$) at $(x^*(\lambda), y^*(\lambda))$, i.e. a minimum that moves with the value of $\lambda$ (see fig. \ref{fig:lagrange_ab}b; cf. eqs. \ref{eq:x_star},\ref{eq:y_star} for $x^*(\lambda), y^*(\lambda))$). Along this "$\lambda$-line", $H''$ vanishes and therefore $H(x^*,y^*,\lambda)=C''(\lambda)$.  Since $C''$ has negative curvature, $H$ has a maximum along the $\lambda$-line at $\lambda^*$ (cf. eq. \ref{eq:lambda*}) and in the three dimensional space $H(x,y,\lambda)$ has a saddle point at $H^*(x^*,y^*,\lambda^*)$. Moreover, the maximum of $H(x^*,y^*,\lambda)$ with respect to $\lambda$ is equal to the minimum of $L(x,y)=H(x,y,\lambda)$ along the line of constraint.

\subsection{Connection to Legendre Transforms}
\label{sec:legendre_transform}
There is a close connection between Lagrange multipliers and Legendre transforms.  To see this, we return to eq. \ref{eq:l+lambda_c} and partition $H$ in yet another way:
\begin{equation}
H(x,y,\lambda)=\bar{L}_1(x,\lambda)+\bar{L}_2(y,\lambda)-\lambda c
\label{eq:lagrange_legendre}
\end{equation}
where
\begin{equation}
\bar{L}_1(x,\lambda)=L_1(x) + m \lambda x = x^2/(2a^2) + m \lambda x,~~~~~~\bar{L}_2(y,\lambda)=L_2(y) + \lambda y = y^2/(2b^2) +\lambda y.
\end{equation}
The Lagrange multiplier conditions for the minimum,
\begin{equation}
(\partial H/\partial x)_{y,\lambda}=(\partial \bar{L}_1/\partial x)_{\lambda}=0 =dL_1/dx + m \lambda,~~~~~~~(\partial H/\partial y)_{x,\lambda}=(\partial \bar{L}_2/\partial y)_{\lambda}=0 = dL_2/dy + \lambda,
\end{equation}
are recognized as the conditions that $\bar{L}_1(m \lambda)$, $\bar{L}_2(\lambda)$ are Legendre transforms of $L_1(x)$, $L_2(y)$, respectively.  

The connection between Lagrange multipliers and the Legendre transform is highlighted by noting that the latter can also be formulated as a duality transformation, mapping convex functions to concave functions and vice versa \cite{arnold} \footnote{\cite{arnold} states that Legendre transforms map convex functions to convex functions; the precise statement depends on the choice of sign in the definition of the Legendre transform}. This mirrors the minimum-maximum relationship of the primary and dual formulations of Lagrange multipliers described above. To keep the main text focused, the definition of Legendre transforms and some of their key properties are developed in Appendix A of the Supplementary Material.
The relationship between Lagrange multipliers and Legendre transforms will play a significant role below.  Virtually all the major equations of analytical mechanics will be shown to arise from Lagrange multipliers, whose dual forms correspond to Legendre transforms in conventional treatments. 

\section{Entropy Maximum Principle}
\label{sec:entropy}
As a first physical application of the dual Lagrange multiplier formalism, we apply it to the entropy maximum principle in thermodynamics \cite{callen,chandler}.  We begin with the primary Lagrange multiplier formalism before proceeding to the dual formalism.

\begin{figure}[h!]
  \begin{center}
\includegraphics[width=16cm]{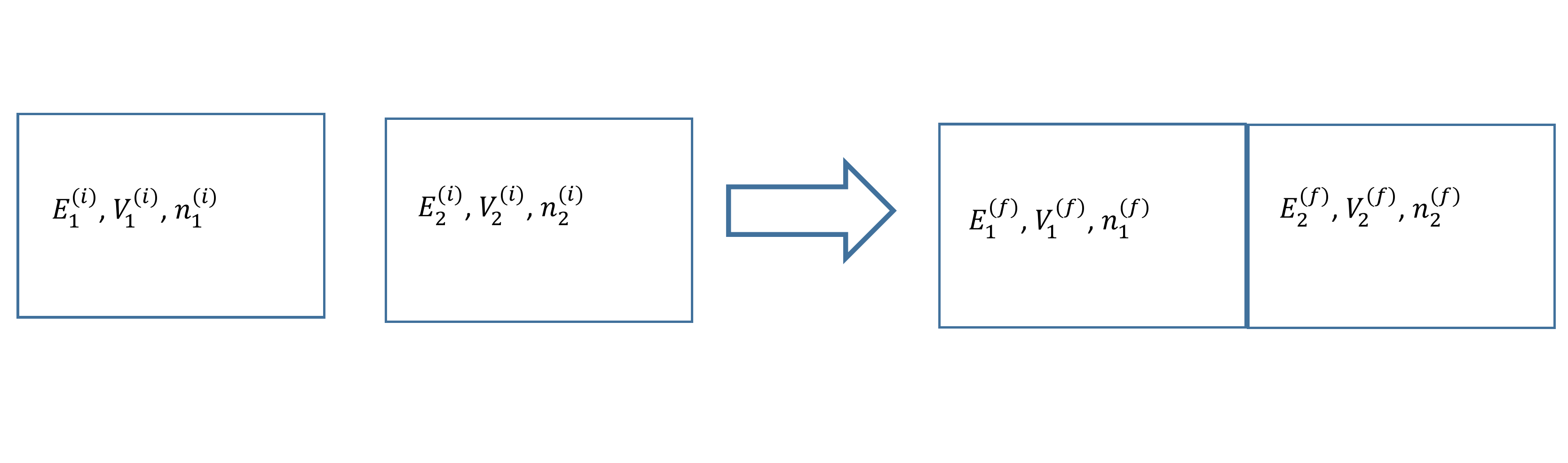}
\end{center}
\caption{\label{fig:composite_system} \footnotesize{Illustration of the entropy maximum principle in thermodynamics. On the left of the arrow is a composite system whose subsystems are not able to exchange $E$, $V$ or $n$. On the right, the same composite system after the two systems are brought into contact and allowed to exchange $E$, $V$ and $n$. (The superscripts $i$ and $f$ stand for initial and final, respectively). The entropy maximum principle answers the question what will be the final values of $E$, $V$ and $n$ for each of the subsystems after they come into contact.}}
\label{fig1}
\end{figure}

\begin{figure}[h]
\begin{center}
\includegraphics[width=12cm]{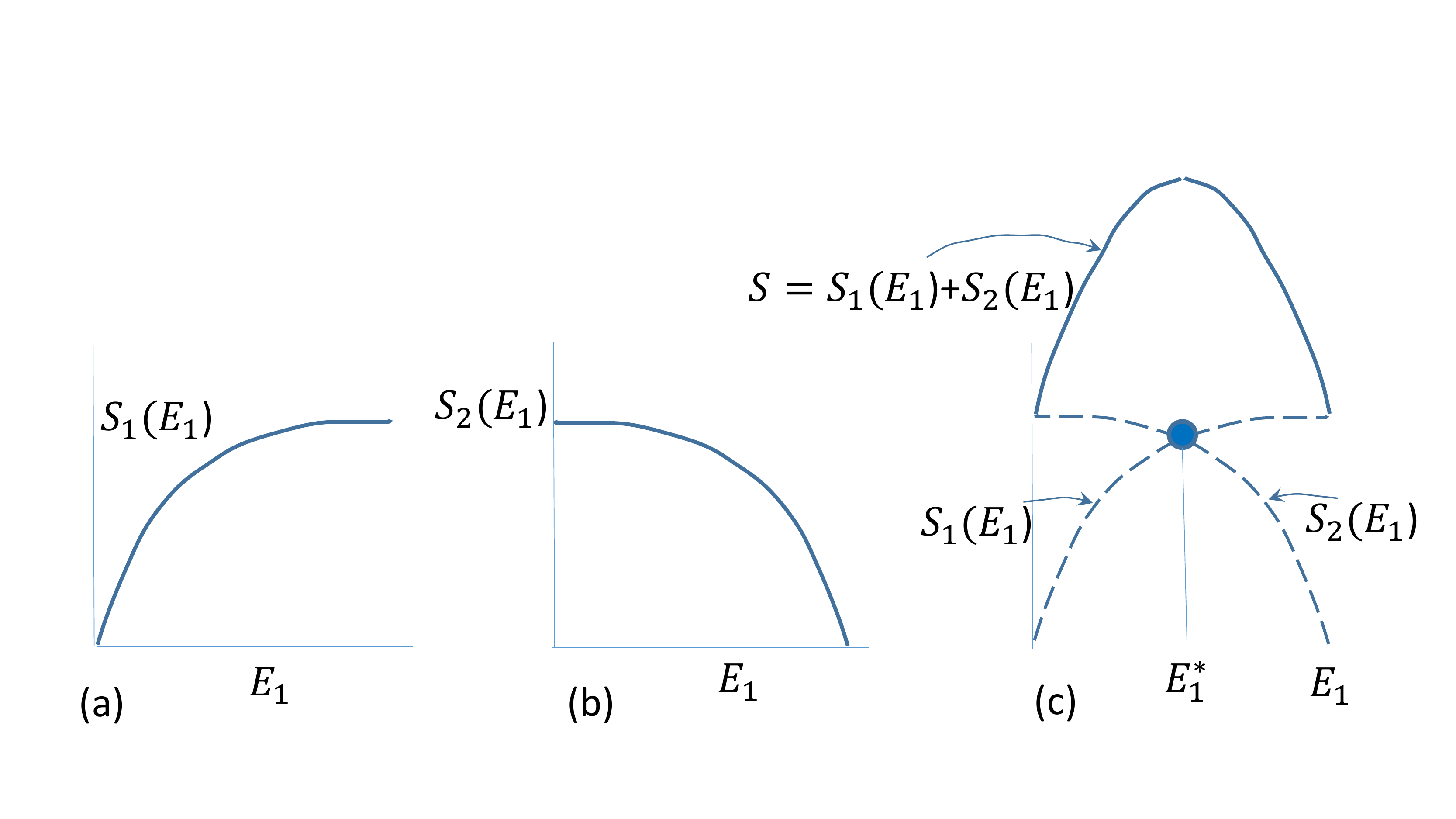}
\end{center}
\caption{\label{fig:S_min}Graphical illustration of the entropy maximum principle.
a)$S_1$ as a function of $E_1$; note that the function is concave.  b) $S_1$ as a function of $E_2$. Since $E_1+E_2=\bar{E}$, when $E_1$ gets larger $E_2$ gets smaller and therefore b) is the reflection of a). c) $S=S_1+S_2$ as a function of $E_1$. Because $S_1$ and $S_2$ are individually concave, their sum is concave and has an interior maximum at $E_1^*$, which determines the value of $E_2^*$, hence the entropy maximum principle determines the partitioning of the total energy $\bar{E}$ between the two subsytems at equilibrium. }
\end{figure}

\subsection{Lagrange multipliers – maximizing the entropy over partitionings of energy}
\label{sec:entropy_lagrange}
Consider two subsystems brought into contact and allowed to exchange energy $E$, volume $V$ and material $n$ (see fig. \ref{fig:composite_system}).  The entropy maximum principle answers the question: what will be the partitioning of $E$, $V$ and $n$ when equilibrium is reached?  For simplicity, we consider that just $E$ can be exchanged between the subsystems. The only input we require is that the entropy $S$ be a concave function of the energy $E$, i.e. $S_1(E_1)$ and $S_2(E_2)$. Define $L = S_1(E_1)+S_2(E_2)$.  The goal is to maximize the combined entropy of the two subsystems, $L$, subject to the constraint that the total energy is conserved, $E_1+E_2=\bar{E}$. To this end, we define $H$:
\begin{equation}
H(E_1,E_2,\beta) = S_1(E_1)+S_2(E_2)-\beta(E_1+E_2-\bar{E}),
\label{eq:h_thermo}
\end{equation}
where we have added the constraint equation with the Lagrange multiplier $\beta$.
Taking derivatives and setting them equal to zero:
\begin{equation}
(\partial H/\partial E_1)_{E_2,\beta}=0 = \partial S_1/\partial E_1-\beta,~~~~~~(\partial H/\partial E_2)_{E_1,\lambda}=0 = \partial S_2/\partial E_2-\beta \Longrightarrow E_2^*(\beta)
\label{eq:e12*}
\end{equation}
\begin{equation}
(\partial H/\partial \beta)_{E_1,E_2} =0 \Longrightarrow E_1+E_2-\bar{E}=0.
\label{eq:thermo_constraint}
\end{equation}
Because of the symmetry of $S_1(E_1)$ and $S_2(E_2)$ we may write eqs. \ref{eq:e12*} in neutral form:
\begin{equation}
\partial S/\partial E-\beta = 0.
\label{eq:e*}
\end{equation}
Equations \ref{eq:e12*} together with eq. \ref{eq:thermo_constraint} provide three equations for three unknowns, $E_1$, $E_2$ and $\beta$:
solving eqs. \ref{eq:e12*} for $E_1^*(\beta)$ and $E_2^*(\beta)$ and substituting into eq. \ref{eq:thermo_constraint} yields 
$\beta^*$. Substituting $\beta^*$ along with $E_1^*(\beta^*)$, $E_2^*(\beta^*)$, $S_1^*(E_1^*(\beta^*))$ and  $S_2^*(E_2^*(\beta^*))$ into eq. \ref{eq:h_thermo} yields $H^*(E_1^*,E_2^*,\beta^*)=L^*(E_1^*(\beta^*),E_2^*(\beta^*))$ where $L^*(E_1^*(\beta^*),E_2^*(\beta^*))$ is the maximum of the constrained problem.

\subsection{Dual formalism – minimizing $\bar{S}$ with respect to inverse temperature}
\label{sec:entropy_dual}
In the dual formulation we substitute eqs. \ref{eq:e12*} for $E_1^*(\beta)$ and $E_2^*(\beta)$ directly into eq. \ref{eq:h_thermo} to obtain:
\begin{equation}
H(E_1^*(\beta),E_2^*(\beta),\beta)=S_1^*(E_1(\beta))+S_2^*(E_2(\beta))-\beta(E_1^*+E_2^*-\bar{E}).
\end{equation}
Calculating $dH(E_1^*(\beta),E_2^*(\beta),\beta)/d\beta=\partial H(E_1^*(\beta),E_2^*(\beta),\beta)/\partial \beta =0$ returns the same value of $\beta^*$ and $H(E_1^*,E_2^*,\beta^*)$ as in the primary procedure. 
Furthermore, 
\begin{equation}
d^2H/d\beta^2 > 0.
\end{equation}
The quantity $\beta$ has the physical interpretation of inverse temperature. To see this, note that at equilibrium the partitioning of energy $E_1^*$, $E_2^*$ between the two subsystems is that which maximizes the entropy. But from eqs. \ref{eq:e12*}, this is just the condition $\partial S_1/\partial E_1 = \partial S_2/\partial E_2$ --- that at equilibrium the temperatures (and hence inverse temperatures) of the two subsystems are equal. Thus, if in the original problem the entropy was concave with respect to energy, in the dual problem $H$ is convex with respect to the inverse temperature $\beta$.

Appendix B of the Supplementary Material discusses the relation of the dual formalism to a Legendre transform of entropy, and Appendix C works out as example for a simple functional form of $S(E)$.

\section{Principle of Least Action}
\label{sec:action}
We now turn our attention to classical mechanics, and take as our starting point the principle of least action.  Inspired by the entropy maximum principle that requires only that the entropy be concave, we assume nothing about the action other than that it is a convex function. Somewhat astonishingly, all of analytical mechanics will emerge, including at the end that the action is the time integral of the Lagrangian.

\subsection{Minimizing the action over partitionings of time}
\label{sec:action_lagrange}
\begin{figure}[h!]
  \begin{center}
\includegraphics[width=12cm]{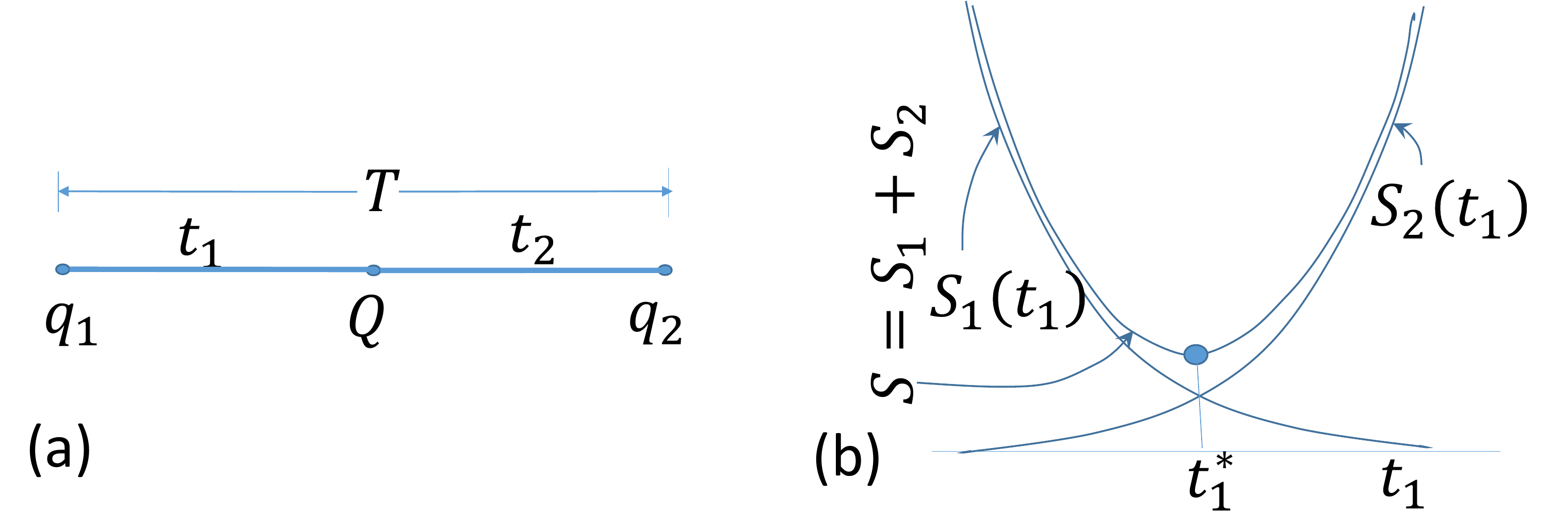}
\end{center}
\caption{a. Formulation of the principle of least action. The endpoints $q_1$, $q_2$ and total time $T$ are fixed.  An intermediate point $Q$ is fixed and we consider all possible partitionings of the total time $T$ such that $t_1+t_2=T$, where $t_1$ is the time from $q_1$ to $Q$ and $t_2$ is the time from $Q$ to $q_2$. b. With each of the two segments in a. is associated an action, $S_1(q_1,Q,t_1)$ and $S_2(Q, q_2, t_2)$ respectively, with the assumption that $S$ is convex with respect to all its arguments and that the action on each segment is minimized. The principle of least action can then be expressed as
$S(q_1,q_2,T) \le S_1(q_1,Q,t_1) + S_2(Q,q_2,t_2)$,
i.e. the composite action $S(q_1,q_2,T)$ minimizes the total action over all partitionings of $t_1+t_2=T$.}
\label{fig:partitioning_t}
\end{figure}

Consider a convex function $S(q_1,q_2,T)$ defined for all paths from $q_1$ to $q_2$ in time $T$.  By analogy with the entropy maximum principle, we partition the path into two segments, $q_1 \rightarrow Q$ and $Q \rightarrow q_2$, and we consider all possible partitionings of time between the two segments subject to the constraint that $t_1+t_2=T$.   We associate with each section an action: $S_1(q_1,Q,t_1)$ and $S_2(Q, q_2, t_2)$ and we assume that the action on each segment is minimized (See Fig. \ref{fig:partitioning_t}).

The principle of least action can then be expressed as:
\begin{equation}
S(q_1,q_2,T) \le S_1(q_1,Q,t_1) + S_2(Q,q_2,t_2).
\end{equation}
To see why this is, note that in the true solution of the problem there is an optimum partitioning of time $T$ into $t_1^*+t_2^*$ so as to minimize $S$ for the entire length; thus any other partitioning of $T$ into $t_1+t_2$ will give a higher value for $S$.  Formulating the problem in terms of Lagrange multipliers we have:
\begin{equation}
\bar{H} = S_1(q_1, Q, t_1) + S_2(Q,q_2,t_2) + E(t_1+t_2-T),
\label{eq:bar_h}
\end{equation}
where $E$ is a Lagrange multiplier for the constraint on total time $T$ and we use a bar over the $H$ to avoid confusion with the Hamiltonian $H$ below.
Taking derivatives with respect to $t_1$, $t_2$ and $E$ and setting them equal to zero we obtain:
\begin{equation}
\partial \bar{H}/\partial t_1 = \partial S_1/\partial t_1 +E=0,~~~~~~\partial \bar{H}/\partial t_2 = \partial S_2/\partial t_2 +E=0
\label{eq:phj1}
\end{equation}
\begin{equation}
\partial \bar{H}/\partial E= t_1+t_2-T=0.
\label{eq:phj3}
\end{equation}
We may write eqs. \ref{eq:phj1} in the neutral form:
\begin{equation}
\partial S/\partial t +E=0.
\label{eq:phj}
\end{equation}
Equation \ref{eq:phj} is well-known in classical mechanics \cite{landau,gutzwiller,heller} and is a precursor of the Hamilton-Jacobi equation (to be obtained below in Section \ref{sec:hj}), with $E$ identified as the energy. Equation \ref{eq:phj3} is of course just the constraint equation. Equations \ref{eq:phj1} also deserve comment: note that the same value of $E$ appears in both equations. By analogy with thermodynamics, where we have seen that the maximum entropy can be expressed as the equality of the inverse temperature of the subsystems, the minimization of the action can be expressed as the equality of the energy of the segments of the trajectory where energy is $-\partial S/\partial t$.

Calculating second derivatives we obtain:
\begin{equation}
\partial^2 S_1/\partial t_1^2 = - \partial E/\partial t_1 >0,~~~~~~\partial^2 S_2/\partial t_2^2 = - \partial E/\partial t_2 >0,
\label{eq:stability1}
\end{equation}
which we may write in the neutral form:
\begin{equation}
\partial^2 S/\partial t^2 = - \partial E/\partial t >0.
\label{eq:stability}
\end{equation}
Equation \ref{eq:stability} is a stability relation for the energy which is a direct result of the convexity of $S$.

\subsection{Dual formalism – maximizing the action with respect to the energy}
\label{sec:action_dual}
Returning to eq. \ref{eq:bar_h}, we can rewrite this equation as:
\begin{equation}
\bar{H}=\bar{S}_1+ \bar{S}_2 - ET,
\end{equation}
where 
\begin{equation}
\bar{S}_1=S_1+Et_1,~~~~~~\bar{S}_2=S_2+Et_2.
\label{eq:bar_s}
\end{equation}
Taking the partial derivative with respect to $E$ and setting it equal to zero we obtain:
\begin{equation}
\partial \bar{S}_1/\partial E-t_1=0,~~~~~~ \partial \bar{S}_2/ \partial E-t_2=0,
\label{eq:dual_hj12}
\end{equation}
or in neutral form
\begin{equation}
\partial \bar{S}/\partial E - t=0.
\label{eq:phje}
\end{equation}
Equation \ref{eq:phje} is also well known in classical mechanics where $\bar{S}(q_1,q_2,E)$ is generally referred to as the abbreviated action \cite{landau,gutzwiller,heller}; we shall also refer to it as the dual action. Note that eqs. \ref{eq:bar_s} may be recognized as a Legendre transform of the action, changing the independent variable from $t$ to $E$. This is a reflection of the close connection between the Lagrange multipliers and Legendre transforms discussed in Section \ref{sec:lagrange_intro}\ref{sec:legendre_transform}. Taking the second derivative with respect to $E$ we find:
\begin{equation}
\partial^2 \bar{S}/\partial E^2 = \partial t/\partial E < 0,
\end{equation} 
i.e. the convexity of $S$ with respect to the partitioning of $T=t_1+t_2$ has an associated dual form of the concavity of $\bar{S}$ with respect to energy.
Appendix D or the Supplementary Material works out in detail an example for a simple functional form of $S(q_1,q_2,t)$.

\subsection{Minimizing the action over partitionings of intermediate position $Q$: generating functions}
\label{sec:action_q}

\begin{figure}[h!]
  \begin{center}
\includegraphics[width=10cm]{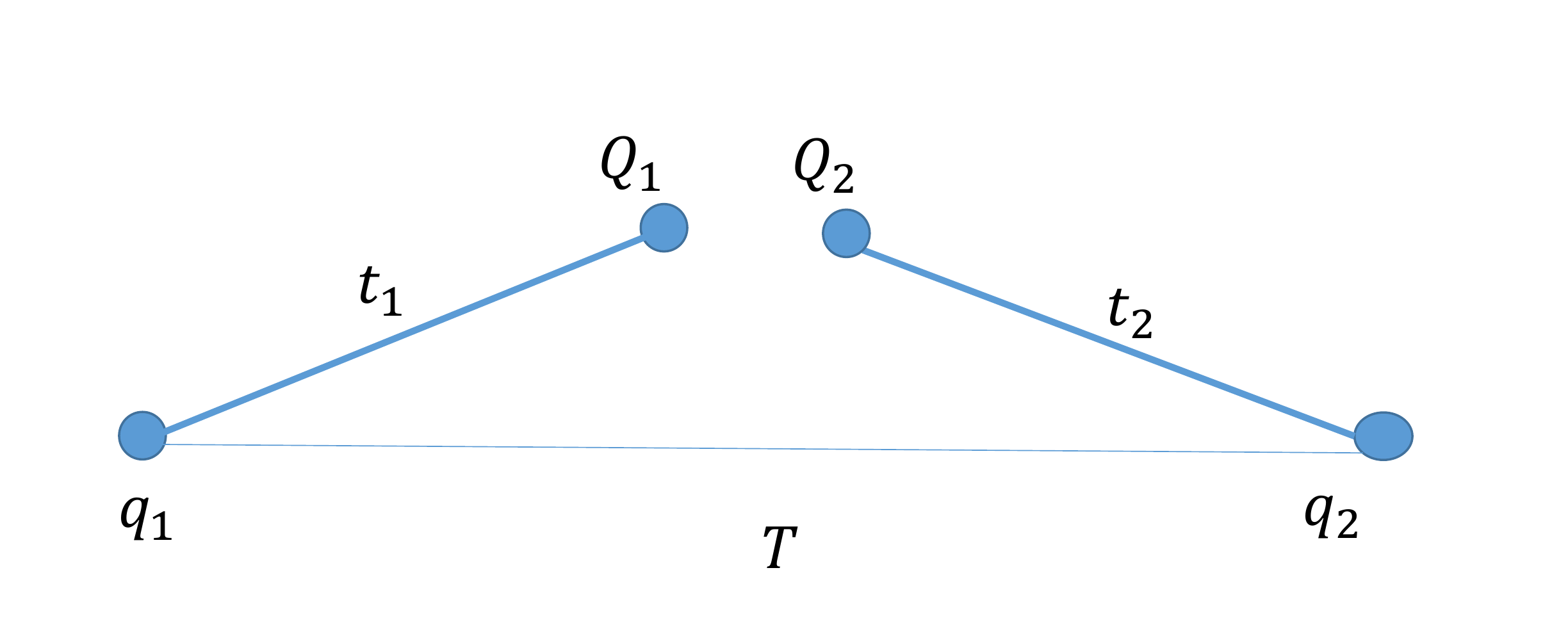}
\end{center}
\caption
{Formulation of the principle of least action, stage 2. The endpoints $q_1$, $q_2$ and total time $T$ are fixed, but the intermediate point $Q$ is now allowed to vary. We consider all possible partitionings of the segment $[q_1,q_2]$ into $[q_1,Q_1]$ and $[Q_2,q_2]$, where $Q_1$ and $Q_2$ need not be identical (see text). We associate with each of the two segments in an action: $S_1(q_1,Q_1,t_1)$ and $S_2(Q_2, q_2, t_2)$ respectively, with the assumption that $S$ is convex with respect to all its arguments and that the action on each segment is minimized. The principle of least action can then be expressed as:
$S(q_1,q_2,T) \le S_1(q_1,Q_1,t_1) + S_2(Q_2,q_2,t_2),$
i.e. the composite action $S(q_1,q_2,T)$ minimizes the total action over all partitionings $t_1+t_2=T$ with the constraint $Q_1=Q_2$ .} 
\label{fig:line_segments_abc}
\end{figure}

We now extend the above treatment of partitioning $T=t_1+t_2$ to allow the endpoints $q_1$ and $q_2$, and the intermediate point $Q$ to vary.  Before proceeding, we consider several possible levels of allowing $Q$ to vary.  The most straightforward is to allow $Q$ to vary but to constrain it to lie along the curve that minimizes $S(q_1,q_2,T)$. A second possibility, one level less constrained, is to allow $Q$ to lie off the curve that minimizes $S(q_1,q_2,T)$.  Finally, we may deconstrain $Q$ even further, by not forcing endpoint $Q_1$ associated with $q_1$ to be equal to endpoint $Q_2$ associated with $q_2$ (see Fig. \ref{fig:line_segments_abc}).  This seems to defy logic, since we must have a continuous curve from $q_1$ to $q_2$, but the apparent inconsistency that $Q_1 \neq Q_2$ presents no problem through the judicious use of Lagrange multipliers. We prefer the latter formulation as the most general.  

Consider again the total action $S$, 
\begin{equation}
S=S_1(q_1,Q_1,t_1) + S_2(Q_2,q_2,t_2)
\end{equation}
but now we specify the coordinate constraints explicitly:
\begin{equation}
t_1+t_2=T, \hspace{1cm} q_1=q’, \hspace{1cm} q_2=q'', \hspace{1cm} Q_2=Q_1. 
\end{equation}
Thus,
\begin{equation}
\bar{S}=S_1(q_1,Q_1,t_1) + S_2(Q_2,q_2,t_2) + E(t_1+t_2-T) - p_2(q_2-q'')+p_1(q_1-q’)+P(Q_2-Q_1).
\end{equation}
We calculate the derivatives of $\bar{S}$ with respect to $q_1$,$q_2$, $Q_1$ and $Q_2$ and obtain:
\begin{equation}
\partial \bar{S}/\partial q_1 =\partial S_1/\partial q_1 + p_1 =0;~~~~~~~\partial \bar{S}/\partial Q_1 = \partial S_1/\partial Q_1 - P=0
\label{eq:f1}
\end{equation}
\begin{equation}
\partial \bar{S}/\partial q_2 =\partial S_2/\partial q_2 - p_2 =0;~~~~~~~\partial \bar{S}/\partial Q_2 = \partial S_2/\partial Q_2 +P=0
\label{eq:f1a}
\end{equation}
Equations \ref{eq:f1} (and similarly eqs. \ref{eq:f1a}) are recognized as the standard equations for the first of four classical generating function (generally referred to as $F_1$) \cite{landau,goldstein,miller,heller}.  Note that as a bonus we get the “handoff condition” \cite{heller},
\begin{equation}
P=-\partial S_2/\partial Q_2 = \partial S_1/\partial Q_1.
\end{equation}

\subsection{Dual formalism – maximizing the action with respect to intermediate momentum $P$}
\label{sec:action_p}
\label{sec:dual_canonical}
Because of the symmetry between eqs. \ref{eq:f1} and \ref{eq:f1a}, we continue the development with just eq. \ref{eq:f1a}. We will come back to eq. \ref{eq:f1} in the Conclusions. To simplify the notation we omit the subscript 2 in eq. \ref{eq:f1a}. 
A variety of alternative forms of the action are possible, corresponding to which terms from the Lagrange multipliers we group with $S(q, Q, t)$. 
\begin{eqnarray} 
  S(q, Q, t) \equiv S_1(q,Q,t)  & \rightarrow & \bar{S}_2(q,P,t) = S_1+PQ_2 \\
			&& \bar{S}_3(p,Q,t) = S_1-pq \\
			&& \bar{S}_4(p_2,P,t_2) = S_1+PQ-pq.
\end{eqnarray}
Taking the various partial derivatives associated with the arguments of $\bar{S}_i,$ $i=2,3,4$, leads to the following set of equations:
\begin{equation}
\partial \bar{S}_2/\partial P = Q, \hspace{1cm} \partial \bar{S}_2/\partial q = p
\label{eq:f2}
\end{equation}
\begin{equation}
\partial \bar{S}_3/\partial Q = -P, \hspace{1cm} \partial \bar{S}_3/\partial p = -q
\label{eq:f3}
\end{equation}
\begin{equation}
\partial \bar{S}_4/\partial P = Q, \hspace{1cm} \partial \bar{S}_4/\partial p= -q,
\label{eq:f4}
\end{equation}
where we have used eqs. \ref{eq:f1}. Equations \ref{eq:f2}-\ref{eq:f4} are recognized as the generating function relations for $F_2$, $F_3$ and $F_4$, respectively \cite{landau,goldstein,miller,heller}.  Again, a close relationship emerges between the Lagrange multiplier method and Legendre transforms: the generating functions $F_2$, $F_3$ and $F_4$ are Legendre transforms of $F_1$.

In each of these cases, we could formulate a dual problem: in principle solve for $t_2^*$, substitute back into the expression for $\bar{S}_i^{(2)}$, and then invoke the relation that the total derivative with respect to the Lagrange multiplier is equal to the partial derivative with respect to the Lagrange multiplier at $t_2^*$.  This procedure turns the original minimization problem turns into a maximation problem with respect to the transformed variable or variables: $p_2$ or $P$ or both (or analogously, $p_1 $ or $P$ or both). 

\subsection{Inverting the functional forms: The Hamilton-Jacobi equation}
\label{sec:hj}
In the theory of canonical transformations, one needs to invert the functional dependence of the generating function relations. We first note that 
the functional dependence $S=S(Q,q,t)$ implies that in the equation $\partial S/\partial t-E=0$, 
\begin{equation}
E=E(Q,q,t).
\label{eq:functional_e}
\end{equation}
Now consider eq. \ref{eq:f1}, $\partial S/\partial q  -   p=0$.  Since $S= S(Q,q,t)$, the functional form of $p$ is $p(Q,q,t)$.  In principle this can be inverted to give $Q=Q(p,q,t)$.  Substituting into eq. \ref{eq:functional_e} gives
\begin{equation}
E(Q,q,t) = E(Q(q,p,t),q,t)=H(p,q,t),
\end{equation}
where we have defined 
\begin{equation}
H(p,q,t)=E(Q,q,t).
\end{equation}
But since 
\begin{equation}
\partial S/\partial q = p
\end{equation}
it follows that 
\begin{equation}
H(p,q,t) = H(\partial S/\partial q,q,t).
\end{equation}
Substituting into eq. \ref{eq:phj1} gives:
\begin{equation}
\partial S/\partial t + H(\partial S/\partial q,q, t)=0,
\label{eq:hj}
\end{equation}
which is the Hamilton-Jacobi equation \cite{landau,goldstein,heller}. The Table
organizes the functional inversions used in this and the following section, from $(q,Q)$ to either $(q,p)$ or $(Q,P)$, i.e. from the natural variables of the primary generating function (two coordinates) to the natural variables of the Hamiltonian (one coordinate and one momentum).

\begin{table}[htb]
\begin{center}
\begin{tabular}{ c c c }
{\bf intrinsic functional dependence} & {\bf primary inversion: $p(q,Q,t) \rightarrow Q(q,p,t)$} & {\bf primary inversion: $P(q,Q,t) \rightarrow q(Q,P,t)$} \\
$S(q,Q,t)$ &	$S(q,Q,t) \rightarrow S(q,Q(q,p,t),t)=S(q,p,t)$ &	$S(q,Q,t) \rightarrow S(q(Q,P,t),Q,t)=S(Q,P,t)$  \\
$E(q,Q,t)$ &	$E(q,Q,t) \rightarrow H(q,Q(q,p,t),t)=H(q,p(q,Q,t),t)$ &	$E(q,Q,t) \rightarrow K(q(Q,P,t),Q,t)=K(Q,P,t)$ \\
$p(q.Q.t)$ &	$p(q,Q,t) \rightarrow Q(q,p,t)$ &	$p(q,Q,t) \rightarrow p(q(Q,P,t),Q,t)=p(Q,P,t)$ \\
$P(q,Q,t)$ &	$P(q,Q,t) \rightarrow P(q,Q(q,p,t),t)=P(q,p,t)$ &	$P(q,Q,t) \rightarrow q(Q,P,t)$ \\
& $(\partial H/\partial q)_Q=(\partial H/\partial q)_p+(\partial H/\partial p)_q (\partial p/\partial q)_Q$ &
$dp/dt=dp(q(Q,P,t),Q,t) = (\partial p/\partial q)_{Q,t}(\partial q/\partial t)_{P,Q} + (\partial p/\partial t)_{q,Q}$ \\
\end{tabular}
\caption{Tabular arrangement of functional inversions required for the development of the theory. }
\end{center}
\end{table}

\subsection{Deriving Hamilton’s Equations of Motion from the Standard and Dual Formulations}
\label{sec:hamilton}

We now proceed to derive Hamilton’s equations of motion \cite{landau,goldstein,miller,heller}.  Consider $S=S(q,Q,t)$ and note that 
\begin{equation}
\partial^2S/\partial q \partial t = \partial^2S/\partial t \partial q.
\label{eq:d2sa}
\end{equation}
Using eqs. \ref{eq:phj} and \ref{eq:f1} we obtain:
\begin{equation}
\partial(-E)/\partial q=\partial p/\partial t,
\end{equation}
which looks very similar to $-\partial H/\partial q=dp/dt$.
Similarly, consider $\bar{S}=\bar{S}(p,P,t)$ and note that
\begin{equation}
\partial^2 \bar{S}/\partial p \partial t = \partial^2 \bar {S}/\partial t \partial p.
\label{eq:d2sb}
\end{equation}
Using eqs. \ref{eq:phj} and \ref{eq:f3} we obtain:
\begin{equation}
\partial E/\partial p = \partial q/\partial t,
\end{equation}
which looks very similar to $\partial H/\partial p = dq/dt$.
Thus we may say, loosely speaking, that Hamilton's two equations of motion emerge from the standard and the dual Lagrange formalism, with each of the two equations arising from a cross-second derivative of $S$, analogous to Maxwell relations in thermodynamics \cite{callen,chandler} .  Hamilton's equations of motion do indeed emerge from these equations, but some additional care is needed as we now show.

Returning to eq. \ref{eq:d2sa}, the LHS gives: 
\begin{equation}
\partial^2S/\partial q \partial t=\partial/\partial q(\partial S/\partial t)= \partial(-E)/\partial q.
\label{eq:d2s_qt}
\end{equation}
Writing the functional dependence of $E$ explicitly:
\begin{equation}
E=E(Q,q,t)=E(Q(q,p,t),q,t)=H(p(Q,q,t),q,t).
\end{equation}
This leads to the relationship:
\begin{equation}
-(\partial E/\partial q)_{Q,t}=
-(\partial H/\partial q)_{p,t} - (\partial H/\partial p)_{q,t}(\partial p/\partial q)_{Q,t}
\label{eq:lhs_a}
\end{equation}
The RHS gives: 
\begin{equation}
\partial^2S/\partial t \partial q= \partial/\partial t(\partial S/\partial q)=\partial p/\partial t,
\label{eq:d2s_tq}
\end{equation}
where $p=p(q,Q,t)$. The key transition to Hamilton's equations is to view $q=q(P,Q,t)$ and $p=(P,Q,t)$, which indicates that the conjugate variables $P,Q$ (which could be e.g. $p(t=0)$, $q(t=0)$) determine $p$ and $q$ at time $t$. Then
\begin{equation}
dp(q(P,Q,t),Q,t)/dt=(\partial p/\partial q)_{Q,t}(\partial q/\partial t)_{P,Q}+(\partial p/\partial t)_{q,Q}, 
\end{equation}
where we identify
\begin{equation}
(\partial q/\partial t)_{P,Q}=dq/dt.
\end{equation}
Therefore,
\begin{equation} 
(\partial p/\partial t)_{q,Q}=dp/dt-(\partial p/\partial q)_{Q,t} dq/dt.
\label{eq:rhs_a}
\end{equation}
Equating eqs. \ref{eq:lhs_a} and \ref{eq:rhs_a} we obtain: 
\begin{equation}
-(\partial H/\partial q)_{p,t}-(\partial H/\partial p)_{q,t} (\partial p/\partial q)_{Q,t} =  dp/dt-(\partial p/\partial q)_{Q,t}dq/dt.
\end{equation}
The factor $(\partial p/\partial q)_{Q,t}$ is common to both sides, so for the two sides to be equal: 
\begin{equation}
-(\partial H/\partial q)_{p ,t}= dp/dt,~~~~~(\partial H/\partial p)_{q,t}=dq/dt.
\end{equation}
Somewhat unexpectedly, both of Hamilton’s equations of motion emerge from eq. \ref{eq:d2s_qt}.

Note that we can also derive Hamilton's equations starting from the dual form $\bar{S}=\bar{S}(p,P,t)$. As before, we begin by noting that 
\begin{equation}
\partial^2 \bar{S}/\partial p \partial t = \partial^2 \bar {S}/\partial t \partial p.
\end{equation}
The LHS gives:
\begin{equation}
\partial^2 \bar{S}/\partial p \partial t=\partial/\partial p(\partial \bar{S}/\partial t) = \partial(-E)/\partial p.
\end{equation}
where $E=E(p,P,t)$. Inverting the functional relationships we obtain:
\begin{equation}
E=E(p,P,t)=E(p,P(q,p,t),t) = H(p,q(p,P,t),t).
\end{equation}
Therefore, the LHS gives:
\begin{equation}
-(\partial E/\partial p)_{P,t}=-(\partial H/\partial p)_{q,t} - (\partial H/\partial q)_{p,t} (\partial q/\partial p)_{P,t}.
\label{eq:lhs_b}
\end{equation}
The RHS gives:
\begin{equation}
\partial^2 \bar{S}/\partial t \partial p = \partial/\partial t(\partial \bar{S}/\partial p)=-\partial q/\partial t.
\label{eq:d2s*_tp}
\end{equation}
where $q=q(p,P,t) \rightarrow P(q,p,t)$. As a result,
\begin{equation}
dq(p(P,Q,t),P,t)/dt = (\partial q/\partial p)_{P,t}(\partial p/\partial t)_{P,Q} + (\partial q/\partial t)_{p,P},
\end{equation}
where we identify $(\partial p/\partial t)_{P,Q}=dp/dt$. Therefore,
\begin{equation}
-\partial q/\partial t = -dq/dt + (\partial q/\partial p)_{P,t}dp/dt.
\label{eq:rhs_b}
\end{equation}
Equating eqs. \ref{eq:lhs_b} and \ref{eq:rhs_b} we obtain again Hamilton's equations of motion:
\begin{equation}
(\partial H/\partial p)_{q,t} = dq/dt, \hspace{1cm} -(\partial H/\partial q)_{p,t}=dp/dt.
\end{equation}

Although both of Hamilton’s equations of motion emerged from eq. \ref{eq:d2sa}, 
$\partial^2S/\partial q \partial t = \partial^2S/\partial t \partial q$, as noted above this equation corresponds more closely to Hamilton’s first equation of motion. 
Similarly, although both of Hamilton’s equations emerged from \ref{eq:d2sb}, $\partial^2 \bar{S}/\partial  p \partial t = \partial^2 \bar{S}/\partial t \partial p$, as noted above this equation corresponds more closely to Hamilton’s second equation of motion.
Thus we may say, loosely speaking, that Hamilton's two equations of motion emerge from the standard and the dual Lagrange formalism, with each of the two equations arising from a cross-second derivative of $S$, analogous to Maxwell relations in thermodynamics \cite{callen,chandler}.  

\subsection{The Lagrangian and the Explicit Form for the Action}
\label{sec:lagrangian}
We now return to $S=S(q,Q,t)$ and substitute the functional form $q=q(Q,P,t)$.  This gives $S(q(Q,P,t),Q,t)$.  As a result, the total derivative of $S$ is given by:
\begin{equation}
dS/dt=(\partial S/\partial q)_{Q,t}(dq/dt)_{Q,P}+(\partial S/\partial t)_{q,Q}
=p \dot{q}-E = L,
\end{equation}
where $L$ is the Lagrangian. Consequently,
\begin{equation}
S=\int_0^T (dS/dt) dt = \int_0^T (p \dot{q} - E) dt= \int_0^T L(q,\dot{q},t) dt,
\label{eq:s_int_l}
\end{equation}
which is the well-known formula expressing the action as the time integral of the Lagrangian \cite{landau,goldstein,heller}.  

We have come full circle.  In standard approaches to analytical mechanics, the starting point is
$S=\int_0^T L(q,\dot{q},t) dt$.  We have arrived at this same formula, not as an axiom, but from an analysis which assumes no more than the convexity of $S$, and hence is essentially just an extension of the variational problem of finding the curve that minimizes the distance between two points. We are amazed.

It is a short step from eq. \ref{eq:s_int_l}, $S=\int_0^T L(q,\dot{q},t)dt$, to the conventional formulation of the principle of least action (Hamilton’s principle).  We derived eq. \ref{eq:s_int_l} based on the partitioning of $S(q_1,q_2,T)$ into two segments and the inequality $S(q_1,q_2,T) \le S_1(q_1,Q,t_1)+S_2(Q,q_2,t_2)$.  Consider now dividing $S(q_1,q_2,T)$ into $N$ segments. Then the inequality takes the form:
\begin{eqnarray}
S(q_1,q_2,T) & \le  S_A(q_1,q_A,t_A)+S_B(q_A,q_B,t_B)+S_C(q_B,q_C,t_C)+ \ldots + S_N(q_{(N)},q_{N+1},t_{N+1})  \nonumber \\ 
& =  \int_{0}^{t_A} L(q,\dot{q}) dt + \int_{t_A}^{t_B} L(q,\dot{q}) dt + \int_{t_B}^{t_C} L(q,\dot{q}) dt + \ldots + \int_{t_N}^{T} L(q,\dot{q}) dt
\end{eqnarray}
Taking the limit $N \rightarrow \infty$ we obtain  
\begin{equation}
S(q_1,q_2,T)= \min \int_{0}^{T} L(q,\dot{q}) dt
\end{equation}
subject to the constraints $q_1=q'=q(t=0)$ and $q_2=q''=q(t=T)$. This is the conventional statement of the principle of least action.

So far, we have implicitly assumed that the $S_1(q_1,Q,t_1)$ and $S_2(Q,q_2,t_2)$ are the same function. But this need not be the case. In fact, if the Hamiltonian is time-dependent, i.e. $H(t_1) \neq H(t_2)$, it implies that $\partial S(q_1,Q,t_1)/\partial t_1 \neq \partial S(Q,q_2,t_2)/\partial t_2$. By allowing the number of partitions in time and distance to go to infinity, and exploiting the full flexibility of $S$, the formalism above extends to time-dependent Hamiltonians.

\subsection{The Canonical 1-form and the Explicit Form for the Abbreviated Action}
\label{sec:pqdot}

In Section \ref{sec:action_dual}, we defined $\bar{S}= S+Et$.
Substituting into eq. \ref{eq:s_int_l}, we obtain 
\begin{equation}
\int (d\bar{S}/dt) dt = \int(p \dot{q} - E) dt+Edt \Longrightarrow \bar{S}=\int pdq,
\end{equation}
the well-known equation for the abbreviated action \cite{landau,goldstein,lanczos,gutzwiller,heller}. 

The abbreviated action satisfies an inequality \cite{tannor_lifeguard}:
\begin{equation}
\bar{S}(q_1,q_2, E) \le \bar{S}(q_1,Q, E) + \bar{S}(Q,q_2, E)
\end{equation}
(see Appendix E of the Supplementary Material).
We can follow the procedure used for $S$ in the previous section, of first partitioning $\bar{S}(q_1,q_2, E)$ into $N$ segments and then letting $N \rightarrow \infty$.  We find
\begin{equation}
\bar{S}(q_1,q_2,E) = \min \int pdq = \min \int 2 dT
\label{eq:maupertuis}
\end{equation}
over all paths from $q_1$ to $q_2$ at fixed energy $E$; now the time for the total path is allowed to vary.   In the second equality in eq. \ref{eq:maupertuis} we have defined the kinetic energy $T=p^2/2m$, $dT=d(p^2/2m)=p dp/m = pdq$. This is the variational equation associated with the names of Maupertuis and Jacobi \cite{landau,goldstein,lanczos,heller}; in the older literature, this is the form that goes by the name of the principle of least action.

\section{Conclusions}
\label{sec:conclusions}

We hope that the reader will share our amazement that essentially all the major formulas of analytical mechanics – the Hamilton-Jacobi equation, generating functions for canonical transformations, Hamilton’s equations of motion, and even the Lagrangian and the action itself -- emerge from just an assumption on the convexity of the action. Hence the entire analysis is essentially just a footnote to the variational problem of finding the curve that minimizes the distance between two points.  

Before addressing the implications, we begin with a review of the main ideas of the paper.  \\
1) First, a doubling of the theory of Lagrange multipliers was introduced, what we call a dual theory. In colloquial language, the dual formalism views the Lagrange multiplier as a way of violating the constraint using some undetermined multiplier. At the end, the violation is corrected by choosing the value of the undetermined multiplier to satisfy the constraint.  For every minimization problem, the dual formalism leads to a maximization problem and vice versa. This duality of Lagrange multipliers is not treated in most of the standard textbooks on analytical mechanics \cite{goldstein,lanczos}, control theory \cite{bryson} or mathematical methods of physics \cite{arfken}, although there are more advanced books and articles that discuss it \cite{hilbert,walsh,bertsekas,kalman}. 

The almost systematic replacement of Legendre transforms by Lagrange multipliers in our treatment hints at a very strong connection between these two.  These connections are explored in detail in a separate publication \cite{twt}. Here we just mention that the definition of Legendre transforms used in \cite{arnold} in which Legendre transforms are cast as a duality transformation, mapping convex functions to concave functions and vice versa (cf. Section A in the Supplementary Material), mirrors the minimum-maximum relationship of the primary and dual formulations of Lagrange multipliers.

2) The dual Legendre formalism was applied first to the entropy maximum principle in thermodynamics and then to the principle of least action in mechanics.  In the former, the total entropy of two subsystems is maximized with the constraint that the total energy is conserved. In the latter, the action $S$ is minimized over two segments of a path connecting $q_1$ to $Q$ and $Q$ to $q_2$, subject to the constraint that $t_1+t_2=T$. The analogy with the entropy maximum principle is clear: in one case we maximize over all partitionings of energy between two subsystems, in the other we minimize the action over all partitionings of time over the two segments of the path. Just as in the former, the only assumption is concavity of the entropy, in the latter the only assumption is convexity of the action – there is no further dynamical input. 

In our treatment, the entropy maximum principle informed the treatment of the principle of least action. But the converse is also true: the principle of least action informs the entropy maximum principle. In discussing the entropy maximum principle, we considered the partitioning of the energy into two subsystems.  By analogy with our discussion of the principle of least action in Section \ref{sec:lagrangian},  if we continue this partitioning process to $N$, and ultimately to an infinite number of partitionings, the entropy maximum principle continues to hold.  This macroscopic statement of the entropy maximum principle in thermodynamics then approaches the microscopic statement of the entropy maximum principle in statistical mechanics, where the maximization is performed over the “partition function”.  We now gain new insight into that term in statistical mechanics, and by the same token we understand the principle of least action as expressed in the form $S= \min \int_{0}^{T} L(q,\dot{q}) dt$ can be viewed as minimizing a partition function expressed as a sum of actions over the partitioning of time into segments.

3) We then continued to apply the dual formalism to mechanics, now allowing the endpoints $q_1$, $q_2$ and the intermediate point $Q$ to vary. Viewing the path of least action from $q_1$ to $q_2$ in time $T$ as finding the shortest distance from $q_1$ to $q_2$ in a Riemannian space, we find three ways to violate the constraints: a) allowing more than the allotted time; b) moving the endpoints closer; c) allowing the end of segment 1, $Q_1$, and the beginning of segment 2, $Q_2$, to not be identical.  The formalism gives as simple consequences: a) the Hamilton-Jacobi equation and its dual form for the abbreviated action $S(E)$; b) the $F_1$ generating function relations for canonical transformations and the $F_2-F_4$ generating functions from dual forms; c) Hamilton's equations of motion. Strikingly, the two Hamilton equations of motion emerge from the standard and the dual Lagrange formalism, with each of the two equations arising from a cross-second derivative of $S$, analogous to Maxwell relations in thermodynamics.  d) the action $S(q_1,q_2,T)$ as the time integral of the Lagrangian and its dual form, $\bar{S}(q_1,q_2,E)$ as the integral of the 1-form $p dq$. 

The skeptical reader may say that the paper merely exchanges the time-tested Legendre transforms in analytical mechanics by Lagrange multipliers, but what has been gained?  To this question there are several compelling answers: \\
\begin{enumerate}
\item In the Lagrange multipler approach, the equations of analytical mechanics come out in a coherent, systematic fashion associated with a hierarchy of derivatives of $S$ (see Appendix F in the Supplementary Material). This is reflected in the striking observation that there are \emph{only derivatives} in this paper ---  no integrals! --- until the Lagrangian appears in the last two subsections of the paper. This is in stark contrast with conventional approaches in which the action $S$ expressed as the time integral of the Lagrangian appears at the very beginning, sometimes as actually the first equation \cite{goldstein,landau,heller}. In fact, it is interesting to note that in conventional approaches (e.g. \cite{landau}) the key equations emerge in almost exactly reverse order (see Appendix G in the Supplementary Material).
\item Since the Lagrange multipliers are each associated with a constraint, the number and character of the Lagrange multipliers is determined a priori; moreover, they have a well-grounded physical basis. In contrast, in conventional treatments the Legendre transforms ($\dot{q} \rightarrow p$, $L \rightarrow H$, $S(t) \rightarrow \bar{S}(E)$, $F_1 \rightarrow F_2-F_4$) appear in a disjointed fashion. Since they are not tied to constraints, when and why they are invoked is not clear. Moreover, the physical basis for the Legendre transformations is not generally transparent. For example, the Legendre transform $\dot{q} \rightarrow p$ leading to $L(q,\dot{q},t) \rightarrow H(q,p,t)$ turns out to be very fruitful but the motivation is not clear a priori. Similarly, in conventional approaches, $F_2-F_4$ emerge as Legendre transforms of $F_1$, but $F_1$ itself is derived almost from thin air, exploiting some hitherto unstated mathematical freedoms.
\item The dual forms of the Lagrange multipliers equations are associated, one-for-one, with what emerge as the dual equations of analytical mechanics. Thus in the approach presented here there is a clear delineation between primary and dual equations.  In contrast, in the conventional treatments the delineation between primary and dual equations is lacking or at best only implied. But more than that: in the Lagrange multiplier formalism a systematic set of maximum principles emerge for the dual equations. These maximum principles could perhaps have been recognized in the Legendre transform formulation but is much clearer in the Lagrange multiplier formulation.
\end{enumerate}







Besides providing a simple and integrated approach to the derivation of the key equations of analytical mechanics, the approach clears up a number of conceptual and notational difficulties in the standard presentations of this material, some of which go back historically to the beginning of the subject. Admittedly there is some subjectivity to these comments, but the author expects that they will resonate with many readers.
\begin{enumerate}
\item In conventional treatments, it is unclear whether the boundary conditions of the Hamilton-Jacobi equation are (in our notation) $(q_1,q_2)$ or $(Q,q_2)$, where $Q$ is arbitrary. It seems this can be traced to the original reformulation of Hamilton's work by Jacobi \cite{nakane} (see Appendix H in the Supplementary Material). Hamilton obtained a pair of partial differential equations (what would later be called Hamilton-Jacobi equations), one for $q_2$ and one for $q_1$.  He seemed to believe that one needed to solve both of these PDEs and then connect the solutions, analogous to the problematic double-ended boundary value problem in the semiclassical propagator (see e.g. \cite{heller}).  Jacobi showed that it was necessary to solve only one of these PDEs, e.g. for $q_2$, and that the other boundary condition could be chosen according to convenience.  This seemingly minor variation on Hamilton's work turned the formalism into a useful calculational tool.  In the present paper, Hamilton's pair of PDEs is reflected in the pair of equations \ref{eq:phj1} (with eq. \ref{eq:hj}) while Jacobi's decoupling of the two equations is reflected in the dichotomous partitioning of the interval $[q_1,q_2]$ into $[q_1,Q_1]$ and $[Q_2,q_2]$, where $Q_1$ and $Q_2$ do not need to be on the line connecting $q_1$ and $q_2$ (cf. fig. 19 in ref. \cite{lanczos}).
\item In conventional treatments, the derivation of the generating functions is mysterious: any function that has a total time derivative can be added to the action without changing the variation. This freedom is exploited to add to the action a function, the number of whose arguments is unclear a priori, which ends up being the central quantity in the theory of canonical transformation, i.e. the generating function which in certain cases is the action itself \cite{goldstein,landau,heller}! In the current approach, the generating functions emerge simply and directly from the Lagrange multiplier formalism. 
\item In conventional treatments, the Hamilton-Jacobi equation is introduced as a way to find the generating function but it is unclear whether the Hamilton-Jacobi equation depends on Hamilton’s equations of motion\footnote{Note that in Jacobi's derivation of the Hamilton-Jacobi equation he seems to have deliberately avoided using Hamilton's equations of motion (see ref. \cite{nakane})}. In the currect approach, the Hamilton-Jacobi equation, as well as the generating functions, emerge \emph{before} Hamilton's equations of motion, a compelling indication that they are independent of Hamilton's equations. 
\item In conventional treatments, there are more than a few ambiguities and inconsistencies in the functional dependence of the central dynamical quantities: the action $S$, the generators $F_1-F_4$, the energy $E$ vs. $H(p,q)$, as well as the coordinate and momentum variables $q$ and $p$: the latter start out being time-independent and only later become time-dependent. Moreover, the total time derivative is used inconsistently in conventional approaches. In the context of $S$, the total time derivative includes the dependence of $S$ on $q_2=q_2(t)$ but not $q_1$. In the context of $q$ and $p$, it is not clear why there should be time-dependence at all. From our approach it is clear when and why $q$ and $p$ are time-dependent, and that the total time derivative indicates that $P$ and $Q$ are fixed. In thermodynamics there has always been great emphasis on the functional dependence of each variable and which variables are held fixed when taking partial derivatives; in this article we have tried to provide the same explicit treatment of the functional dependences of the dynamical variables in analytical mechanics leading to a rigorous formulation of the partial and total derivatives.
\end{enumerate}

We close with some comments about the larger implications of this work. Note that the only assumption in our formulation is that there is a convex function $S(q_1,q_2,t)$.  There is no mention of kinetic or potential energy; there is no a priori assumption about energy, momentum, Hamiltonian or Lagrangian, these quantities emerging almost from thin air. Since convexity is a property of the shortest distance between two points, it would appear that this paper is simply an extension of that problem to general convex functions $S(q_1,q_2,t)$, with the physics entering through the choice of $S(q_1,q_2,t)$\footnote{It is intriguing to note that despite the collective historical perception of Hamilton's contributions to dynamics, this was precisely the goal he had in his seminal work: to find a function $S(q_1,q_2,t)$ that would allow the solution of the dynamical problem just by differentiations and substitutions \cite{nakane}}. Seen purely from a mathematical perspective, the variational problem of the shortest distance between two points takes on an intriguing variation if one adds an intermediate point $Q$ and reexpresses the problem in terms of $q_1$, $q_2$ and $Q$. Thus, it seems that much of analytical mechanics is essentially just a footnote to the mathematical problem of finding the shortest distance between two points. As such, this work should lead to a critical reevaluation of the place of analytical mechanics within the mathematics and physics literature.

\acknow{Financial support for this work came from the Israel Science Foundation (1094/16 and 1404/21), the German-Israeli Foundation for Scientific Research and Development (GIF) and the historic generosity of the Harold Perlman family. The author is grateful to Dr. Dahvyd Wing for helpful discussions throughout this work.}

\showacknow{} 

\bibliography{refs}

\end{document}


\maketitle
\thispagestyle{firststyle}
\ifthenelse{\boolean{shortarticle}}{\ifthenelse{\boolean{singlecolumn}}{\abscontentformatted}{\abscontent}}{}

\section*{Appendix A: Dual Formulation of Legendre Transforms}

In this Appendix, we first review the definition of Legendre transforms. We follow the presentation in ref \cite{arnold}, which is not the standard definition in physics but which is remarkably close in spirit to the approach to Lagrange multipliers in this paper. In fact, we will show an explicit connection between the two. 

We next consider derivatives of the Legendre transform. We calculate the first derivative of the Legendre transform in two ways, each providing a somewhat different perspective. We then turn to the second derivative and show that the Legendre transform of a convex function is a concave function, and vice versa. In ref \cite{arnold}  this result is expressed as that the Legendre transform of a convex function is a convex function, but the sign in the definition of the Legendre transform is reversed from the definition used in thermodynamics and mechanics, and that adopted here.

\begin{figure}[h!]
  \begin{center}
\includegraphics[width=12cm]{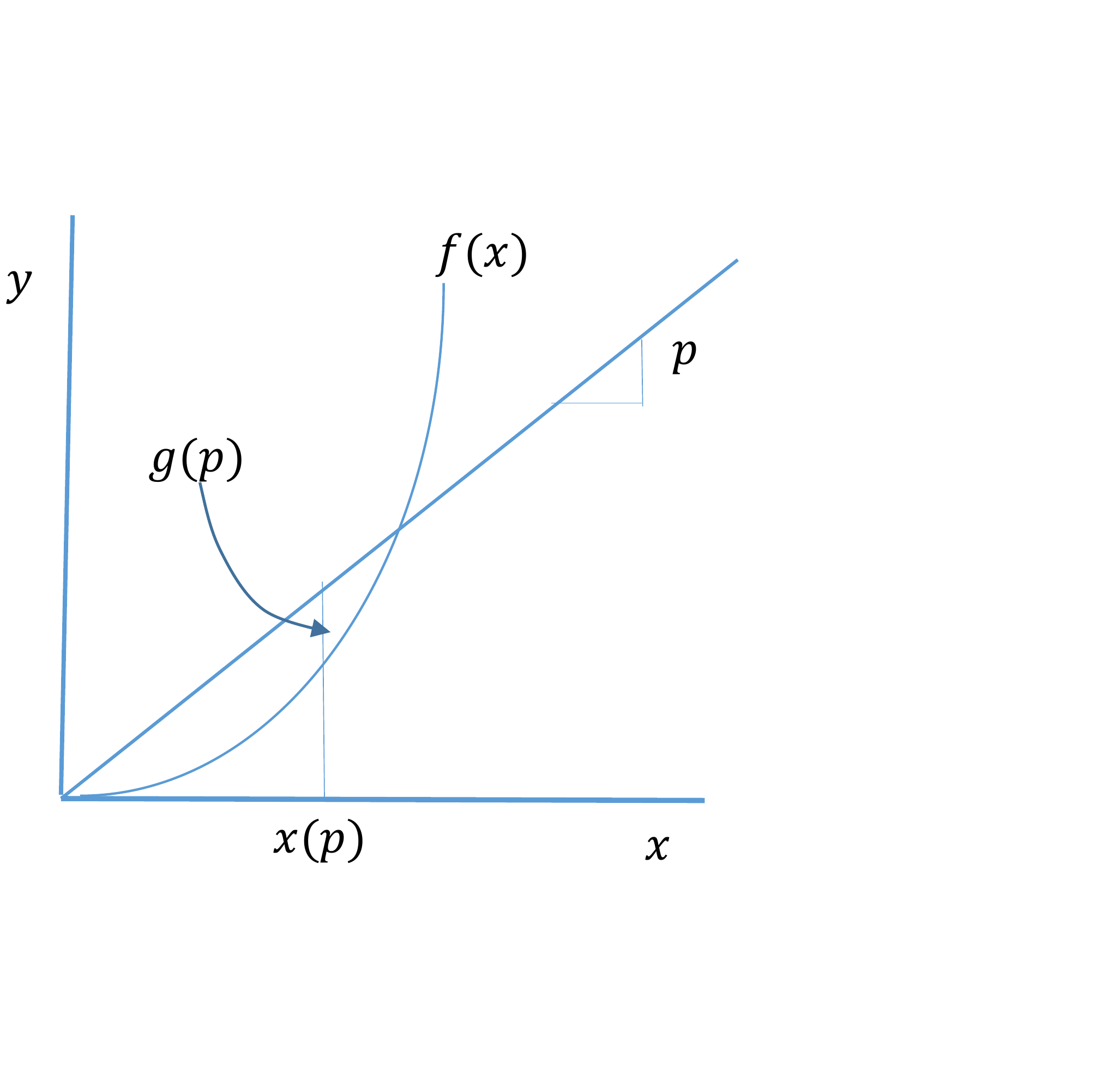}
\end{center}
\caption{Graphical representation of the Legendre transform, following ref. \cite{arnold}. For an explanation see the text.}
\label{fig:legendre}
\end{figure}

We begin with the defintion of the Legendre transform. Consider a function $y=f(x)$, where $f(x)$ is a convex function, $f''(x) >0$. The Legendre transform of the function $f$ is a new function $g$ of a new variable $p$, which is constructed in the following way. We draw the graph of $f$ in the $x,y$ plane. Let $p$ be a given number and consider the straight line $y=px$. We take the point $x=x(p)$ at which the curve is furthest from the straight line in the vertical direction: for each $p$ the function 
\begin{equation} 
F(p,x)=px-f(x)
\end{equation}
has a maximum with respect to $x$ at the point $x(p)$, call it $x^*$ (see fig. \ref{fig:legendre}). The point $x^*(p)$ is defined by the extremal condition 
\begin{equation}
\partial F/\partial x=0=p-f'(x^*),
\end{equation}
or 
\begin{equation}
df/dx^*=p.
\label{eq:legendre_deriv1}
\end{equation}
The Legendre transform is defined as 
\begin{equation}
g(p)=-F(p,x^*(p))=-(px^*-f(x^*)).
\end{equation}
Differentiating $g(p)$ we obtain 
\begin{equation}
dg/dp=-x^*.
\label{eq:legendre_deriv2}
\end{equation}
Comparing eqs. \ref{eq:legendre_deriv1} and \ref{eq:legendre_deriv2} we see the Legendre reciprocity relationship for first derivatives.

It is useful to define the function $h(p)=x^*$. It follows that $h$ satisfies
\begin{equation}
f'(h(p))=p~~~~{\rm or}~~~~ h=(f')^{-1}
\label{eq:legendre_deriv} 
\end{equation}
In terms of this definition, 
\begin{equation}
g(p)=-(px^*-f(x^*))=-(ph(p)-f(h(p))).
\end{equation} 
Differentiating $g(p)$ we find:
\begin{equation}
dg/dp=-(h(p)+(p-f'(h(p)))dh(p)/dp)=-h(p),
\end{equation}
where we have used eq. \ref{eq:legendre_deriv}. This establishes a connection betwen $dg/dp$ and $h=(f')^{-1}$, i.e. the derivative of the transform $g(p)$ is equal up to a sign to the inverse of the derivative of the original function $f(x)$.

Turning to the second derivatives, Legendre transforms take convex functions to concave functions and vice versa. To see this, we begin by reviewing the rule for differentiating an inverse function. Let $y=f(x)$, $f^{-1}(y)=x$. Using the property that $(dx/dy)(dy/dx)=1$ we have $dx/dy=1/(dy/dx)$. Identifying $dy/dx$ with $f'(x)$ and $dx/dy$ with $(f^{-1})'$ and choosing $y=a$ we have
\begin{equation}
(f^{-1})'(a)=1/(f'(f^{-1}(a))).
\end{equation}
Replacing $f$ with $f'$ we obtain:
\begin{equation}
(f'^{-1})'(a)=1/(f''(f'^{-1}(a))).
\label{eq:deriv2_legendre}
\end{equation}
Applying eq. \ref{eq:deriv2_legendre} to the second derivative of $g$ with respect to $p$ and setting $a=p$ we obtain
\begin{equation}
g''(p)=d^2g/dp^2=-dh(p)/dp=-d(f')^{-1}/dp=-1/f''(f'^{-1}(p))=-1/f''(h(p))=-1/f''(x) < 0
\label{eq:convexity_proof}
\end{equation}
where the final inequality follows since $f''(x) >0$.  This shows that the Legendre transform, with the physicist's definition, takes convex functions to concave functions and vice versa.

In the text, all our cases using Lagrange multipliers involve linear constraints. Under these conditions, the addition of the constraint using a Lagrange multiplier becomes equivalent to a Legendre transform.  As such, the dual form of the Lagrange multiplier has the same property as the dual form of the Legendre transform, i.e. it takes convex functions to concave and vice versa.  If the form of the constraint equation is nonlinear there will be local relations of the type of \ref{eq:convexity_proof}, but they will not be global.

\section*{Appendix B.  The Entropy Maximum Principle: Relationship to Legendre Transforms}
\label{sec:entropy_legendre}
In the text, we had the equation
\begin{equation}
H(E_1,E_2,\beta) = S_1(E_1)+S_2(E_2)-\beta(E_1+E_2-\bar{E}),
\label{eq:h_thermo}
\end{equation}
where we added the constraint equation with the Lagrange multiplier $\beta$.
Because of the symmetry of eq. \ref{eq:h_thermo} for $H$ with respect to $E_1$ and $E_2$, there is an instructive alternative way to develop the dual formalism.
We rewrite eq. \ref{eq:h_thermo} as follows:
\begin{equation}
H(E_1,E_2,\beta) = \bar{S}_1(E_1)+\bar{S}_2(E_2)+\beta E,
\label{eq:h_thermo_symmetric}
\end{equation}
where $\bar{S}_1(E_1)=S_1(E_1)-\beta E_1$ and $\bar{S_2}(E_2)=S_2(E_2)-\beta E_2$. Taking the derivatives with respect to $E_1$ and $E_2$ and setting them equal to zero we obtain:
\begin{equation}
(\partial H/\partial E_1)_{E_2,\beta}=0 = \partial \bar{S}_1/\partial E_1-\beta,~~~~~~(\partial H/\partial E_2)_{E_1,\lambda}=0 = \partial \bar{S}_2/\partial E_2-\beta.
\label{eq:h_thermo_deriv_symmetric}
\end{equation}
Because of the symmetry of $\bar{S}_1(E_1)$ and $\bar{S}_2(E_2)$ we consider just a single equation and write it in neutral form, 
\begin{equation}
\bar{S}(E)=S-\beta E,~~~~~~\partial \bar{S}/\partial E-\beta = 0.
\label{eq:entropy_legendre}
\end{equation}
From the second of these equations, we find $E^*(\beta)$ and $S(E^*(\beta))$.
Substituting into eq. \ref{eq:entropy_legendre} we find 
\begin{equation}
\bar{S}(\beta)=S(E^*(\beta))-\beta E^*(\beta).
\end{equation}
Equations \ref{eq:entropy_legendre} are the defining relation for a Legendre transform $S(E) \rightarrow \bar{S}(\beta)$.
Then
\begin{equation}
d\bar{S}/d\beta = \partial \bar{S}/\partial \beta = -E^*;
\end{equation}
comparing with eq. \ref{eq:entropy_legendre} we see that the Legendre reciprocity relationship for first derivatives is satisfied.  Calculating the second derivative with respect to 
$\beta$,
\begin{equation}
d^2\bar{S}/d\beta^2 >0.
\end{equation}
There are two ways to understand the sign of the second derivative.  
The first is to notice that in eq. \ref{eq:entropy_legendre}, $\beta$ can still be viewed as a Lagrange multiplier appending to $S$ the constraint $E=0$. The actual value of the constant of constraint cannot change the duality relationship and as such if $S$ is concave with respect to $E$, $\bar{S}$ must be convex with respect to $\beta$.  The second way is to recall that Legendre transforms map convex functions to concave functions and vice versa (see Appendix A) and thus if $d^2S/dE^2 <0$ then $d^2\bar{S}/d\beta^2 >0$. 

\section*{Appendix C: A Simple Example of the Entropy Maximum Principle}
As a simple example of the entropy maximum principle, let $E(S)=S^2$, implying that $S(E)=E^{1/2}$. Then $L = S_1(E_1)+S_2(E_2)= E_1^{1/2}+E_2^{1/2}$ and
\begin{equation}
H(E_1,E_2,\beta)=(E_1^{1/2}+E_2^{1/2})-\beta(E_1+E_2-\bar{E}).
\label{eq:entropy_example_h}
\end{equation}
Taking derivatives and setting them equal to zero:
\begin{equation}
(\partial H/\partial E_1)_{E_2,\beta}=0 = (1/2) E_1^{-1/2}-\beta \Longrightarrow E_1^*=1/(4 \beta^2)
\end{equation}
\begin{equation}
(\partial H/\partial E_2)_{E_1,\lambda}=0 = (1/2) E_2^{-1/2}-\beta \Longrightarrow E_2^*=1/(4 \beta^2)
\end{equation}
Substituting the values of $E_1^*$ and $E_2^*$ back into the constraint equation and solving for $\beta$ we find:
\begin{equation}
E_1^*+E_2^*-\bar{E}=0 =1/(2 \beta^2)-\bar{E} \Longrightarrow \beta^*=(1/(2\bar{E}))^{1/2}.
\label{eq:beta*}
\end{equation}
Substituting $E_1^*$, $E_2^*$ and $\beta^*$ back into eq. \ref{eq:entropy_example_h} gives
\begin{equation}
H^*(E_1^*,E_2^*,\beta^*)=(2\bar(E))^{1/2}.
\end{equation}
In the dual formulation we substitute  $E_1^*$, $E_2^*$ and $\beta$ into $H$ to obtain:
\begin{equation}
H(E_1^*,E_2^*,\beta)=1/(2 \beta) + \beta \bar{E}=H(\beta).
\end{equation}
Taking the total derivative with respect to $\beta$ and setting it equal to zero we find:
\begin{equation}
dH/d \beta=0 \Longrightarrow 1/(2 \beta^{*2})=\bar{E},
\end{equation}
consistent with eq. \ref{eq:beta*} for $\beta^*$. Furthermore,
\begin{equation}
d^2H/d \beta^2 = \beta^{-3} >0;
\end{equation}
in constrast with the original problem in which the entropy was concave with respect to energy, in the dual problem $H$ is convex with respect to $\beta$. Thus, in the original problem $L=S_1(E_1)+S_2(E_2)$ is a maximum with respect to the partitioning of $\bar{E}$, i.e. $L^*(E_1^*,E_2^*)$, while in the dual problem $H(E_1^*,E_2^*,\beta)$ is a minimum with respect to $\beta$ at $H^*(E_1^*,E_2^*,\beta^*)=L^*(E_1^*,E_2^*)=(2\bar{E})^{1/2}$.

\section*{Appendix D. The Principle of Least Action: A Simple Example}
As an example of the principle of least action, consider 
\begin{equation}
S(q_1,q_2,T)=m(q_2-q_1)^2/(2T),
\label{eq:s_example}
\end{equation}
\begin{equation}
S_1(q_1,Q,t_1)=m(Q-q_1)^2/(2 t_1)~~~~~~S_2(Q,q_2,t_2)=m(q_2-Q)^2/(2t_2).
\end{equation}
Note that $S$ as well as $S_1$ and $S_2$ are convex with respect to all their arguments, e.g.
\begin{equation}
\partial^2 S_1/\partial t_1^2 = m(Q-q_1)^2/t_1^3 > 0.
\end{equation}
We define
\begin{equation}
\bar{H} = \bar{S}_1 + \bar{S}_2 - ET
\label{eq:bar_s_example}
\end{equation}
where
\begin{equation}
\bar{S}_1 = m(Q-q_1)^2/(2 t_1) + Et_1,~~~~\bar{S}_2 = m(q_2-Q)^2/(2 t_2) + Et_2.
\label{eq:fp_bar_s}
\end{equation}
Then
\begin{equation}
\partial \bar{S}_1/\partial t_1 = -m(Q-q_1)^2/(2 t_1^2) + E = 0 \Longrightarrow t_1^*(E) =(m/2E)^{1/2}(Q-q_1). 
\label{eq:fp_t*}
\end{equation}
and similarly
\begin{equation}
t_2^*(E) =(m/2E)^{1/2}(q_2-Q). \label{eq:fp_t2*}
\end{equation}
Substituting $t_1^*(E)$ and $t_2^*(E)$ back into the constraint equation $t_1+t_2-T=0$ we may solve for $E^*$:
\begin{equation}
t_1^*+t_2^*=T=(m/2E)^{1/2}(q_2-q_1) \Longrightarrow E^*=m(q_2-q_1)^2/(2T^2)
\label{eq:e*1}
\end{equation}
Substituting $t_1^*$, $t_2^*$ and $E^*$ back into eq. \ref{eq:bar_s_example} we obtain
\begin{equation}
\bar{H}^*(t_1^*,t_2^*,E^*)=(2m)^{1/2}({m(q_2-q_1)^2}{2T})^{1/2}(q_2-q_1)-m(q_2-q_1)^2T/(2T^2) = m(q_2-q_1)^2/(2T)=S(q_1,q_2,T).
\end{equation}
We have returned to eq. \ref{eq:s_example}, exactly where we started, but note the significance of what we have found: the optimum partitioning of $T$ into $t_1^*$ and $t_2^*$ returns $S_1(t_1^*(E^*))+S_2^*(t_2^*(E^*))=S^*(t_1^*,t_2^*,E^*)$, confirming that $S(q_1,q_2,T) \le S_1(Q,q_1,t_1)+S_2(q_2,Q,t_2)$, i.e. the equality corresponds to the partitioning of time that minimizes the total action and the inequality corresponds to all other partitionings.

We now turn to the dual formulation. Substituting $t_1^*$ and $t_2^*$ into $\bar{S}_1$ and $\bar{S}_2$, respectively, defines 
the abbreviated or dual form for the action:
\begin{equation}
\bar{S}_1(Q,q_1,E)=(2mE)^{1/2}(Q-q_1)~~~~~~\bar{S}_2(q_2,Q,E)=(2mE)^{1/2}(q_2-Q).
\label{eq:bar_s_e}
\end{equation}
Taking the first derivative with respect to $E$ we find
\begin{equation}
d \bar{S}_1/dE= (m/2E)^{1/2}(Q-q_1)=t_1^*(E),~~~~~~d \bar{S}_2/dE= (m/2E)^{1/2}(q_2-Q)=t_2^*(E)
\label{eq:d_barS_dE}
\end{equation}
consistent with eqs. \ref{eq:fp_t*} and \ref{eq:fp_t2*}.
Taking the second derivative we find that
\begin{equation}
\partial^2 \bar{S}_1/\partial E^2 = -1/2 (m/2)^{1/2}E^{-3/2}(Q-q_1)=\partial t^*/\partial E < 0
\end{equation}
and similarly for $\partial^2 \bar{S}_2/\partial E^2$;
hence the dual problem is concave whereas the original problem was convex.
Substituting eqs. \ref{eq:bar_s_e} back into eq. \ref{eq:bar_s_example} we obtain:
\begin{equation}
\bar{H}=(2mE)^{1/2}(Q-q_1)+(2m/E)^{1/2}(q_2-Q)-ET=(2mE)^{1/2}(q_2-q_1)-ET.
\end{equation}
Then
\begin{equation}
d\bar{H}/dE=0 \Longrightarrow T=(m/2E^*)^{1/2} (q_2-q_1) \Longrightarrow E^*=m(q_2-q_1)^2/(2T^2),
\end{equation}
consistent with eq. \ref{eq:e*1} for $E^*$. Finally, we calculate $\bar{H}^*(t_1^*(E^*),t_2^*(E^*),E^*)$:
\begin{equation}
\bar{H}^*(t_1^*(E^*),t_2^*(E^*),E^*)=(2m)^{1/2}(m(q_2-q_1)^2/(2T^2))^{1/2}(q_2-q_1)-m(q_2-q_1)^2T/(2T^2)=m(q_2-q_1)^2/(2T)=S(q_1,q_2,T).
\end{equation}
Again we have returned exactly to where we started, eq. \ref{eq:s_example}, but now this value of $\bar{H}(t_1^*(E),t_2^*(E),E)$ is a \emph{maximum} with respect to the Lagrange multiplier $E$ 
(this is clear from the conditions $d\bar{H}/dE=0$ and $d^2\bar{H}/dE^2 = d^2\bar{S}_1/dE^2+d^2\bar{S}_2/dE^2 < 0$).

We can use this example to reinforce an important point about the connection between Lagrange multipliers and Legendre transforms.
From eqs. \ref{eq:fp_bar_s}-\ref{eq:fp_t*}, $\bar{S}_1$ and $\bar{S}_2$ are recognized as Legendre transforms of $S_1$ and $S_2$. The Legendre transform maps convex functions to concave functions and vice versa, mirroring the minimum-maximum relationship of the primary and dual formulations of Lagrange multipliers. However, since the Legendre transform contains no information on the \emph{constant} of the constraint, in this case $T$,  there is no way of calculating $E{^*}$ and hence the actual value of $\bar{H}(t_1^*,t_2^*,E^*)$ without information on the constant $T$.  In other words the convex $\rightarrow$ concave duality is not sufficient for finding the actual value of the constrained minimum. For the latter, one must substitute $S_1(Q,q_1,E)$ and $S_2(q_2,Q,E)$ back into eq.\ref{eq:bar_s_example} for $\bar{H}$, where the constant of constraint $T$ does appear. Maximizing with respect to $E$ and substituting $E^*$ back into $\bar{H}$ one obtains equality of the minimum of the primary problem and the maximum of the dual problem.

\section*{Appendix E: The Principle of Least Action: Feynman's Lifeguard Problem}

\begin{figure}[h!]
  \begin{center}
\includegraphics[width=20cm]{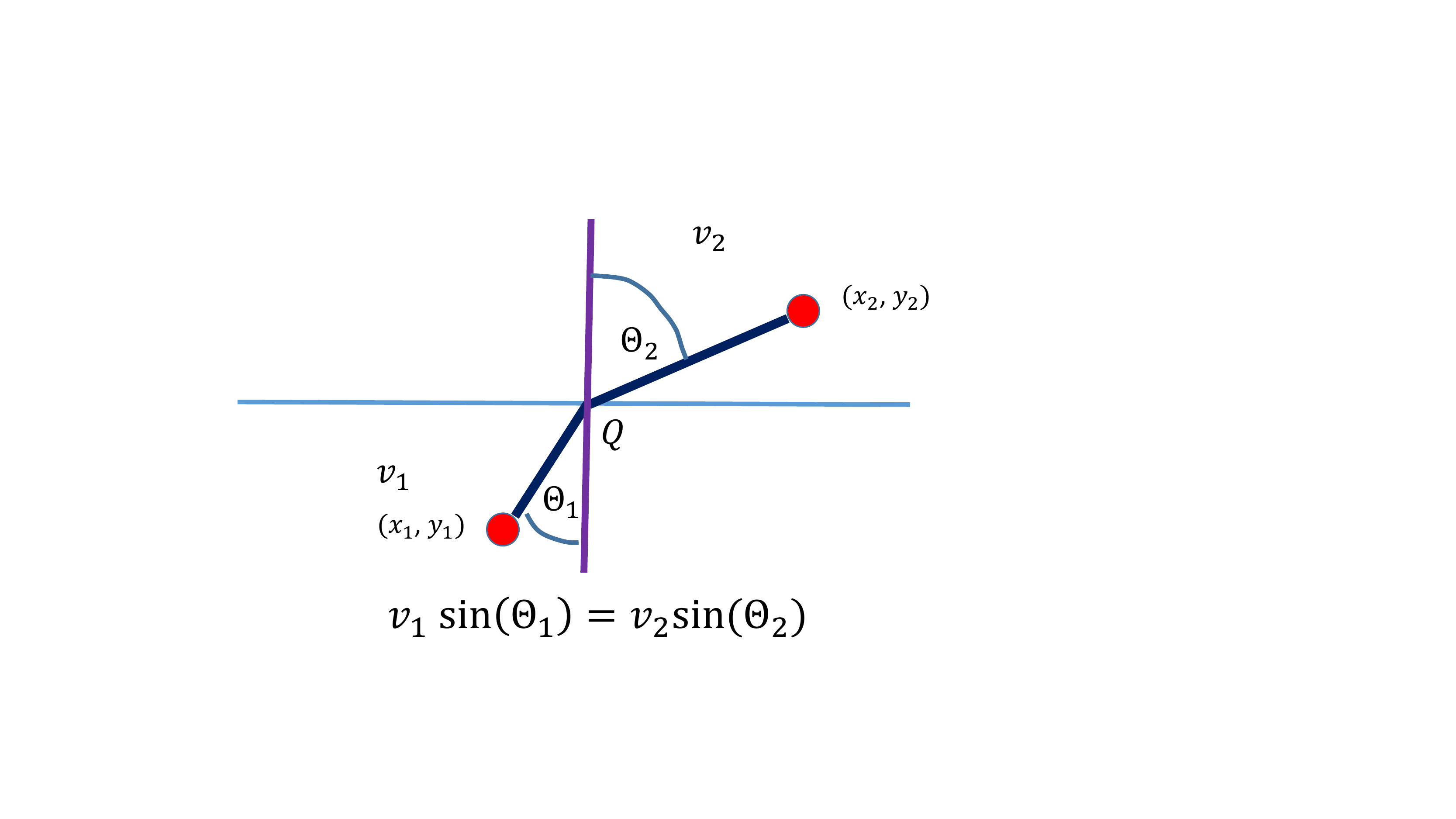}
\end{center}
\caption{Graphical representation of the lifeguard problem.  For light, according to Snell's law, if $v_1 > v_2$ this implies that $\theta_1 > \theta_2$. For matter, satisfying the principle of least action there is an inverted Snell's law: if $p_1 > p_2$ and hence $v_1 > v_2$ this implies that $\theta_1 < \theta_2$ as shown in the figure.}
\label{fig:lifeguard}
\end{figure}

Consider a system with $V_1(x,y)=0,~~~~y<0$, $V_2=V,~~~y\ge 0$; Consider an initial point is $q_1=(x_1,y_1),~~~y_1<0$ and a final point $q_2=(x_2,y_2),~~~y_2 >0$. Assume that the velociies in the different regions are $\dot{q}_1$ and $\dot{q}_2$ respectively.  The problem is to find the point $Q=(x,y=0)$ that minimizes the time to go from $q_1$ to $q_2$ (see Fig. \ref{fig:lifeguard}). This problem was used to beautiful effect by Feynman in volume I of the Feynman Lectures on Physics to introduce Fermat's principle of least time in optics \cite{feynman}. Feynman shows that the relationship of the angles $\theta_1$ and $\theta_2$ is given by Snell's law, $n_1 \theta_1=n_2 \theta_2$where $n$ is the index of refraction, $n=c/v$. To give the problem a human flourish, Feynman shows the formal equivalence with the problem of a lifeguard who sees a person drowning and has to choose the path of least time given a velocity $v_1$ that he can run on the sand and a velocity $v_2$ that he can swim in the ocean. After starting with this example, Feynman goes on to develop all of geometrical optics from this principle of least time.
The Feynman lifeguard problem is arguably the simplest non-trivial example of a minimum principle in more than one dimension. It also very close in spirit to the approach in this paper in that it is dichotomous: $q_1$ and $q_2$ are given and the object is not to solve a full variational problem to minimize the desired quantity but to minimize with respect to one intermediate point. Essentially, there is an implicit assumption that once the intermediate point $Q$ is chosen, the straight lines from $q_1$ to $Q$ and from $Q$ to $q_2$ minimize the time. Put another way, if instead of just two values of the velocity the problem was defined in terms of a velocity that changed continuously with $y$, one would require a full-fledged variational treatment. If one were to discretize the variational problem into $N$ discrete stages and minimized simultaneously with respect to all $N$ stages, one would in fact obtain Euler's (as opposed to Lagrange's) approach to the calculus of variations \cite{lanczos}. The point of this discussion is that the 2-stage process is not only well-suited for pedagogical purposes, but in fact in some sense represents the fundamental building block whose extension to $N$-stages leads to a full variational treatment.  

In this Appendix we will revisit Feynman's lifeguard problem, but our interest in not in the trajectories of light but of particles. Instead of Fermat's principle of least time we will study the same problem using the principle of least action. To keep the terminology clear we will distinguish between the long action, $S(q_1,q_2,T)$ and the short action $S(q_1,q_2,E)$, and we will examine the minimzation of both of these quantities. The long action and the short action are related by a Legendre transform from $t$ to $E$ \cite{gutzwiller}. In the literature, the minimization of the long action is generally called Hamilton's principle (Feynman calls it the principle of least action), while minimization of the short action is generally called Mauperuis's principle of least action (it sometimes goes by the name of Lagrange or Jacobi, and historically is just called the principle of least action).  To highlight the symmetry in the terminology and to avoid misunderstanding we will call these as the principle of least long action and the principle of least short action.  The principle of least long action yields a trajectory that has both a coordinate space orbit and a time-schedule: we know where the trajectory is along the orbit at every instant of time. In contrast, the principle of least short action yields just a coordinate space orbit with no time schedule. Furthermore, in the former the total time is fixed but the energy is not. In the latter, the energy is fixed but the total time is not. What is true, but not at all obvious, is that the coordinate space orbit obtained by the principle of least long action is the same as the coordinate space orbit obtained by the principle of least short action. Note that this statement is trivial in 1-d, since in 1-d the coordinate orbit is just a straight line, which could have been deduced without any dynamical information. Thus the need to study at least a 2-d system.  As mentioned above, the Feynman lifeguard problem is arguably the simplest such system. In this Appendix we study the principle of long action and the principle of short action for the Feynman lifeguard problem and show explicitly the equivalence of the orbits.  

As mentioned above, although the Feynman lifeguard problem is only a 2-stage problem, it is the fundamental building block of the general variational problem. As such, relationships between $S(q_1,q_2,t)$ and $\bar{S}(q_1,q_2,E)$ that emerge in the context of this simple example will generally apply to the general case.  In particular, the fact that the principle of least long action provides a minimum with respect to the orbit implies that $S(q_1,q_2,t)$ is convex with respect to $q$; similarly, the fact that the principle of least short action provides a minimum with respect to the orbit implies that $\bar{S}(q_1,q_2,E)$ is convex with respect to $q$. Since $S(q_1,q_2,t)$ and $\bar{S}(q-1,q_2,E)$ are related by a Legendre transformation from $t$ to $E$, $q$ is a passive variable in the Legendre transform.  The existence of two principles of least action with the same orbit for both implies that the convexity of the passive variable is unchanged under a Legendre transform. More generally, the relationship of short action $\bar{S}(q_1,q_2,E)$ to $S(q_1,q_2,t)$ should be understood in the larger context of the duality transforms discussed in the text, where here a new aspect enters, namely the role of the passive variable. in duality.

Before proceeding to the lifeguard problem, we pause to make contact with the conventional literature.  
\begin{equation}
S(q_1,q_2,t) = \int_0^T L(q,\dot{q},t) dt;~~~~\bar{S}(q_1,q_2,E)=\int_{q_1}^{q_2} pdq.
\end{equation}
In the common situation that $L=T-V$, $E=T+V$, $T=p^2/(2m)=m\dot{q}^2/2$
\begin{equation}
S(q_1,q_2,t) = \int_0^T T(q,\dot{q}) - V(q,t) dt;~~~~\bar{S}(q_1,q_2,E)=\int_{q_1}^{q_2} (2m(E-V)^{1/2} dq.
\end{equation}
Writing $p=m\dot{q}$ and $dq=dq/dt dt = \dot{q} dt$, $\bar{S}(q_1,q_2,E)$ is sometimes written in the bastardized form 
\begin{equation}
\bar{S}(q_1,q_2,E)=\int_{q_1}^{q_2} m\dot{q}dq = \int_{t_1}^{t_2} m\dot{q}^2 dt = \int 2 T dt;
\end{equation}
we say bastardized because as explained above the principle of least short action is a statement about the coordinate dependence of the orbit without any reference to schedule. Moreover, in the short action the energy is fixed but the end points in time are not, which is inconsistent with writing the limits of the integral as $t_1$ and $t_2$. (The usual explanation for this inonsistency is that the latter are not given a priori as in the principle of least long action, but should be understood, along with $t$, as parametrizing the coordinate space orbit that minimizes the short action). For the lifeguard problem, $L_1=1/2 m \dot{x}_1^2 + 1/2 m\dot{x}_2^2$, $L_2=1/2 m \dot{x}_1^2 + 1/2 m\dot{x}_2^2-V$. The integral expression $S(q_1,q_2,t) = \int_0^T L(q,\dot{q},t) dt$, which will appear only at the end of our presentation, was introduced here in order to make plausible the forms of $S_1$ and in particular $S_2$ below.

Turning to the lifeguard problem, we take $S_1$ and $S_2$ to be: 
\begin{equation}
S_1(q',q'',T) = m/2 [(x-x_1)^2/t_1 + y_1^2/t_1];~~~~~~S_2=m/2[(x_2-x)^2/t_2 + y_2^2/t_2]-Vt_2
\end{equation}
Both $S_1$ and $S_2$ are seen to be convex with respect to $t_1$, $t_2$ and $x$. As in the main text, we will first minimize $S_1$ and $S_2$ with respect to $t_1$ and $t_2$ respectively, subject to the constraint that $t_1+t_2=T$ for fixed $Q$; following this we will minimize with respect to $Q$. Thus we write
\begin{equation}
\bar{S}(q_1,q_2,t)=m/2((x-x_1)^2+y_1^2)/t_1+m/2((x_2-x)^2+y_2^2)/t_2-Vt_2 +E(t_1-t_2-T).
\label{eq:long_action_lifeguard}
\end{equation}
\begin{equation}
\partial \bar{S}/\partial t_1=-m/2((x-x_1)^2+y_1^2)/t_1^2+E=0;~~~~~~\partial \bar{S}/\partial t_2 = -m/2((x-x_2)^2+y_2^2/t_2^2-V+E=0.
\label{bar_s_supp}
\end{equation}
Solving for $t_1^*$ and $t_2^*$ we find:
\begin{equation}
t_1^*=(m/2E)^{1/2}((x-x_1)^2+y_1^2)^{1/2};~~~~~~t_2^*=(m/(2(E-V)))^{1/2}((x_2-x)^2+y_2^2)^{1/2}. 
\end{equation}
Substituting $t_1^*$ and $t_2^*$ back into eq. \ref{eq:long_action_lifeguard} for $\bar{S}$, and simplifying we obtain:
\begin{equation}
\bar{S}=(2mE)^{1/2}((x-x_1)^2+y_1^2)^{1/2}+(2m(E-V))^{1/2}((x_2-x)^2+y_2^2)^{1/2}.
\end{equation}
Minimizing with respect to $x$ we find:
\begin{equation}
\partial \bar{S}/\partial x=0=(2mE)^{1/2} (x-x_1)/((x-x_1)^2+y_1^2)^{1/2}+(2m(E-V))^{1/2}(x-x_2)/((x_2-x)^2+y_2^2)^{1/2}.
\label{eq:snell_long_action}
\end{equation}
Noting that 
\begin{equation}
(x-x_1)/((x-x_1)^2+y_1^2)^{1/2}=\sin \theta_1;~~~~~~(x-x_2)/((x_2-x)^2+y_2^2)^{1/2}=-\sin \theta_2
\end{equation}
we find
\begin{equation}
E^{1/2} \sin \theta_1 = (E-V)^{1/2} \sin \theta_2.
\end{equation}
This is a Snell's law for trajectories, with the roles of $n_1$ and $n_2$ played by $E^{1/2}$ and $(E-V)^{1/2}$ respectively.  In fact, it is an inverted Snell's law, in the sense that the angle $\theta_1 < \theta_2$ if $p_1 > p_2$ and hence if $v_1 > v_2$. The correctness of this result can be checked by noting that when a trajectory reaches $y=0$ its $x$-velocity doesn't change, and its $y$ velocity changes to satisfy conservation of energy $1/2 m v_1^2 = E$, $1/2 m v_2^2 + V=E$.  

We now turn to the principle of least short action. The expression $\int_{q_1}^{q_2} pdq$ becomes, for the 2-stage process,
$p_1 \Delta q_1 + p_2 \Delta q_2$. Since $p_1=(2mE)^{1/2}$,$p_2=(2m(E-V))^{1/2}$, $\Delta q_1=((x_1-x)^2+y_1^2)^{1/2}$;~~$\Delta q_2=((x_2-x)^2+y_2^2)^{1/2}$
the task is to minimize 
\begin{equation}
\bar{S}=(2mE)^{1/2}((x_1-x)^2+y_1^2)^{1/2}+(2m(E-V))^{1/2}((x_2-x)^2+y_2^2)^{1/2}.
\label{eq:short_action_lifeguard}
\end{equation}
But this is identical to eq. \ref{eq:snell_long_action}, i.e. the trajectory of least shsort action will be the same as the trajectory for the least long action (for a fixed total time $T$). From this follow a number of conclusions. 1) Since any spatially varying potential can be made up by concatenating lifeguard problems, the trajectory of least short and least long action will be the same for all potentials.  2) Equation \ref{eq:long_action_lifeguard} can be written as
\begin{equation}
S(q_1,q_2,T) \le S_1(q_1,Q,t_1) + S_2(Q,q_2,t_2)
\label{eq:long_action_variational}
\end{equation}
subject to $t_1+t_2=T$, while eq. \ref{eq:short_action_lifeguard} can be written as
\begin{equation}
\bar{S}(q_1,q_2, E) \le \bar{S}(q_1,Q, E) + \bar{S}(Q,q_2, E). 
\label{eq:short_action_variational}
\end{equation}
Thus eq. \ref{eq:short_action_variational} follows from eq. \ref{eq:long_action_variational} for the lifeguard problem (cf. eq. \ref{eq:snell_long_action}).
But since the dynamics in an arbitrary potential can be described by concatenating lifeguard problems, this implies that in general the variational principle for the long action
\begin{equation}
S(q_1,q_2,T) = \min \int_0^T L(q,\dot{q},t) dt
\end{equation}
implies the variational principle for the short action
\begin{equation}
\bar{S}(q_1,q_2,E)=\min \int_{q_1}^{q_2} pdq.
\end{equation}
See Sections 4G-H in the main text for further discussion.

\newpage
\section*{Appendix F: Tabular Arrangement of the Hierarchy of Derivatives of $S$}

\begin{figure}[thp]
  \begin{center}
\includegraphics[width=20cm]{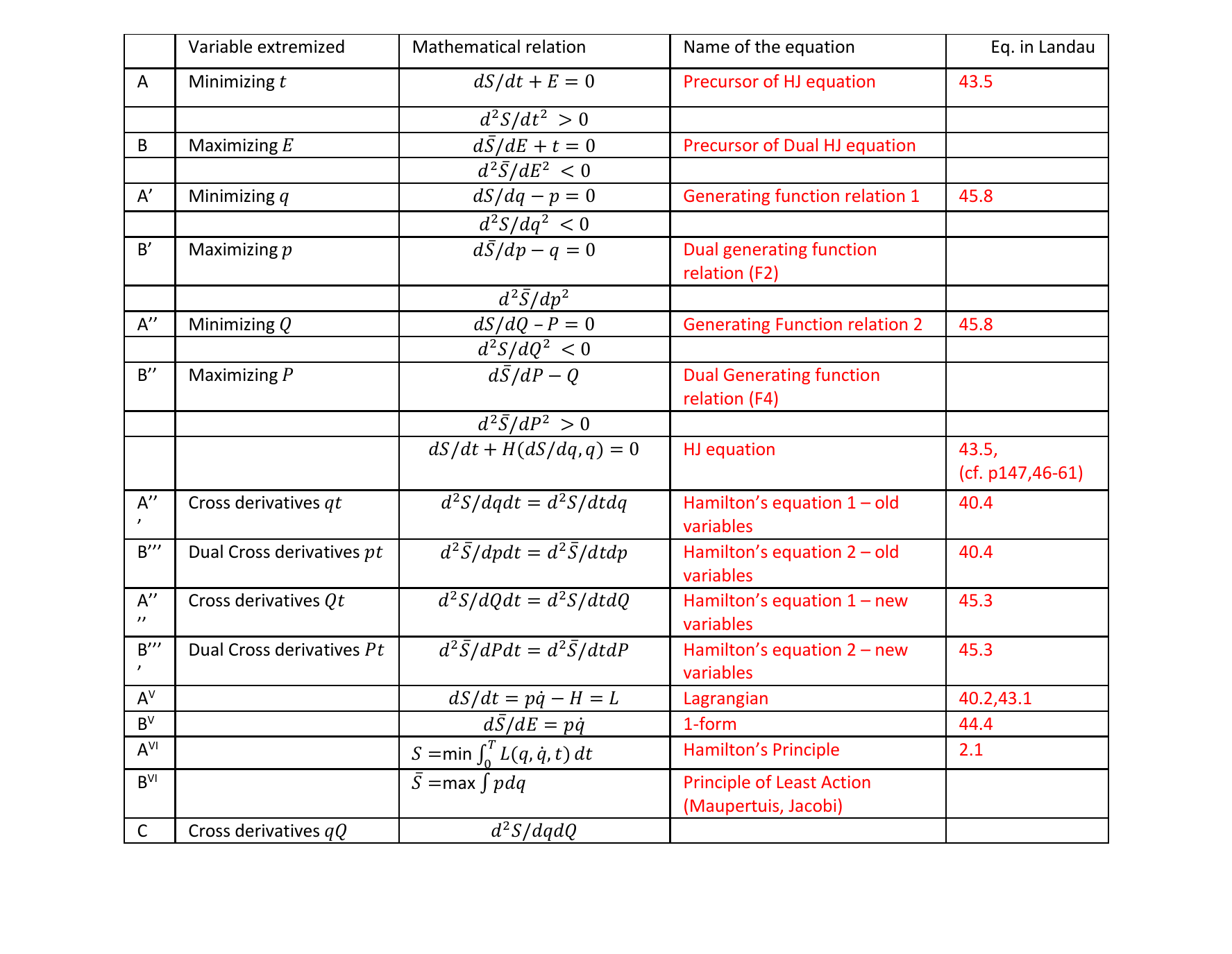}
\end{center}
\caption{Tabular arrangement of the derivatives of $S$ in the dual approach and the name of the equation. The Table highlights how in the dual approach the key equations of analytical mechanics are associated with a hierarchy of the action $S$, its Legendre transform $\bar{S}$, and their first, second and cross derivatives. Where applicable the name of the resulting equation is given.}
\label{fig:tabular_equations}
\end{figure}
\newpage
\section*{Appendix G: Tabular Arrangement of the Equations of Analytical Mechanics in the Conventional and the Current Approach}

\begin{figure}[thp]
  \begin{center}
\includegraphics[width=20cm]{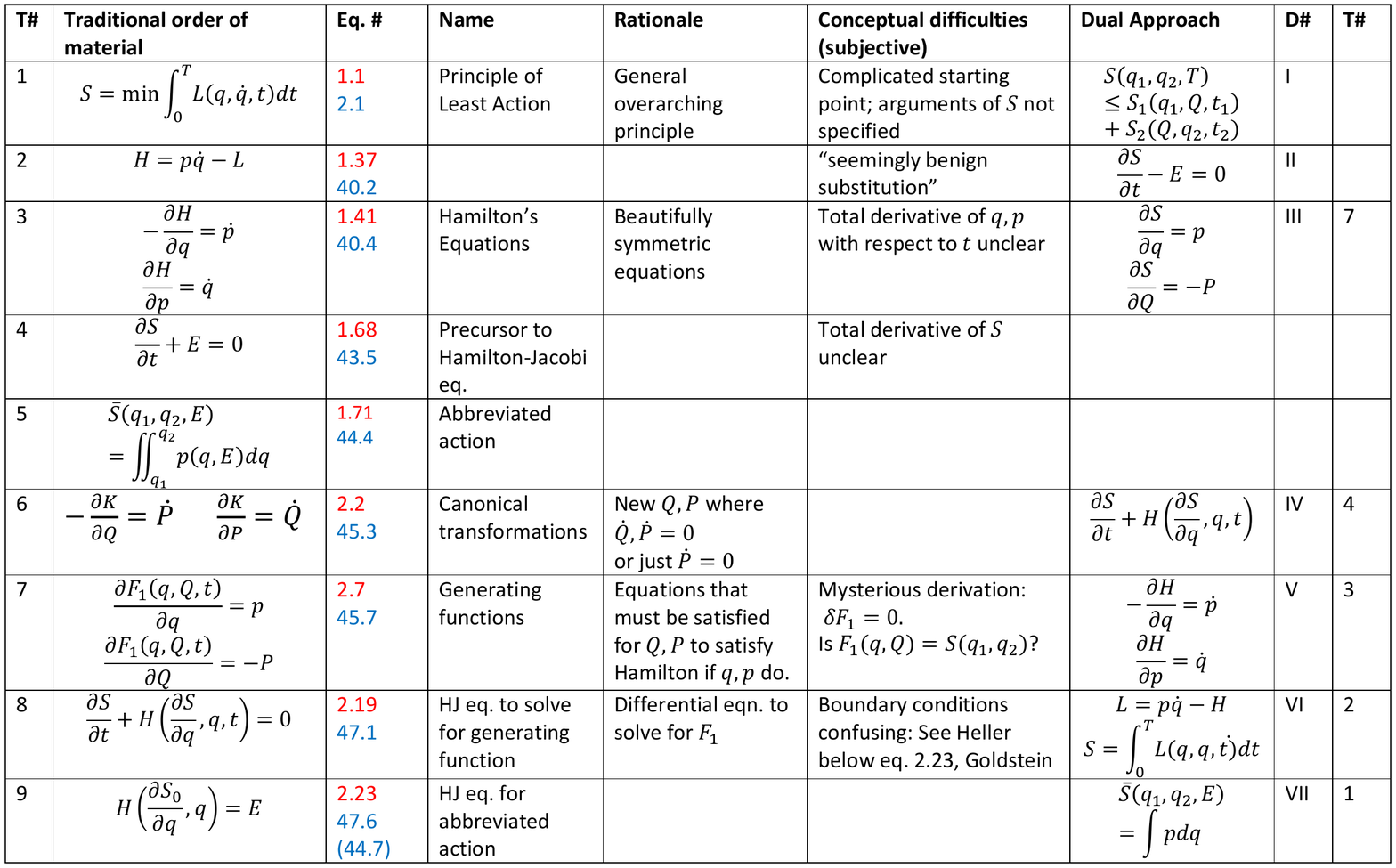}
\end{center}
\caption{Tabular arrangement of the equations of analytical mechanics in the conventional and the current approach. T\# indicates the sequence of equations in the traditional approach and D\# indicates the sequence of equations in the dual approach. It is striking that the key equations of analytical mechanics come out almost systematically in reverse order in the dual approach compared with the conventional approach. This can be seen clearly in the rightmost column. The equation numbers in the traditional approach refer to refs. \cite{heller} and \cite{landau} respectively.} 
\label{fig:tabular_equations}
\end{figure}

\section*{Appendix H: A Brief History of the Hamilton-Jacobi Equation} It is interesting to review the historical development of the ideas of analytical mechanics from the perspective of the derivation presented here. We start with a disclaimer that we are no experts on the history, but base our remarks on the clear and extensive summary of the oriiginal papers of Hamilton and Jacobi given in \cite{nakane}. The fact that there are only derivatives in the paper, i.e. no integrals, until the Lagrangian appears in the last two subsections of the paper, is consistent with Hamilton's vision: a formulation of mechanics, analogous to his formulation of optics, in which all trajectory information could be obtained by derivatives and substitutions \cite{nakane}. The integral equation expressing the principle of least action (Hamilton's principle) as the time integral of the Lagrangian was derived by Hamilton in his original paper as almost an afterthought. In fact, it was Hamilton who first defined what is today called the Lagrangian in this same afterthought, consistent with the sequence of derivations as they emerge in this paper. Historically, Jacobi adding three insights to Hamilton's formulation: 1) that the Hamiltonian could be time-dependent; 2) Hamilton had a derived two coupled "Hamilton-Jacobi" equations, one for  $S(Q,q_2,t_2)$ and one for $S(q_1,Q,t_1)$. Jacobi realized that these equations could be decoupled, freeing up a double ended boundary problem to have just a single fixed boundary. Our claim is that the insight of Jacobi, in the language of this paper, was that one can introduce an intermediate point $Q$ that is not on the path of minimal $S$ from $q_1$ to $q_2$; once $Q$ does not need to lie on the path of minimal $S$ from $q_1$ to $q_2$, then $Q$ and its conjugate momentum $P$ can be complicated functions of $q$ and $p$ and do not need to have the conventional interpretation of coordinate and momentum. 
These developments, beginning with Jacobi, were taken further a few years later by Delaunay, who performed a series of hundreds of canonical transformations in his work on celestial mechanics \cite{nakane,lanczos,gutzwiller}. The modern formulation of canonical transformations using generating functions apparently appears in the work of Poincare, possibly for the first time, who refers to them as Jacobi's equations \cite{lanczos}. Thus it seems that it was only in the course of the 60 years following Hamilton that the full machinery of canonical transformation as we know it today was developed. 



%

\bibliography{refs}